\documentclass{aa}  

\usepackage{graphicx}
\usepackage{txfonts}
\usepackage{amstext}

\begin{document} 

   \title{Parsec-scale jets driven by high-mass young stellar objects.}
\subtitle{Connecting the au- and the parsec-scale jet in IRAS\,13481-6124\thanks{Based on observations collected at the European Southern Observatory La Silla, Chile, 087.C-0951(A), 087.C-0951(B), 087.C-0489(C), and 090.C-0371(C).}}

   \author{R. Fedriani
          \inst{1,2}
          \and
          A. Caratti o Garatti\inst{1}
          \and D. Coffey\inst{1,2}
          \and R. Garcia Lopez\inst{1}
          \and S. Kraus\inst{3}
          \and G. Weigelt\inst{4}
          \and B. Stecklum\inst{5}
          \and T.P. Ray\inst{1}
          \and C. M. Walmsley\fnmsep\thanks{This work is dedicated to the memory of C. M. Walmsley. We would love to acknowledge Malcolm's dedication, passion, kind explanations, and fruitful conversations. Without these, this paper would not be the same.
Thank you, Malcolm.}
          }

   \institute{Dublin Institute for Advanced Studies, School of Cosmic Physics, Astronomy \& Astrophysics Section, 31 Fitzwilliam Place, Dublin 2, Ireland\\
              \email{fedriani@cp.dias.ie}
         \and
             University College Dublin, School of Physics, Belfield, Dublin 4, Ireland
          \and 
             School of Physics, Astrophysics Group, University of Exeter, Stocker Road, Exeter EX4 4WL, UK
          \and
             Max-Planck-Institut f\"ur Radioastronomie, Auf dem H\"ugel 69, D-53121 Bonn, Germany
          \and
             Th\"uringer Landessternwarte Tautenburg, Sternwarte 5, 07778 Tautenburg, Germany
             }

   \date{Received 27 October 2017 / Accepted 23 May 2018}

 
  \abstract
%
   {Protostellar jets in high-mass young stellar objects (HMYSOs) play a key role in the understanding of star formation and provide us with an excellent tool to study fundamental properties of HMYSOs.}
   {We aim at studying the physical and kinematic properties of the near-IR (NIR) jet of IRAS\,13481-6124 from au to parsec scales.}
   {Our study includes NIR data from the Very Large Telescope instruments SINFONI, CRIRES, and ISAAC. Information about the source and its immediate environment is retrieved with SINFONI. The technique of spectro-astrometry is performed with CRIRES
to study the jet on au scales. The parsec-scale jet and its kinematic and dynamic properties are investigated using ISAAC.}
   {The SINFONI spectra in the $H$ and $K$ band are rich in emission lines that are mainly associated with ejection and accretion processes. Spectro-astrometry is applied to the Br$\gamma$ line, and for the first time, to the Br$\alpha$ line, revealing their jet origin with milliarcsecond-scale photocentre displacements ($11-15$\,au). This allows us to constrain the kinematics of the au-scale jet and to derive its position angle ($\sim216\degr$). ISAAC spectroscopy reveals H$_2$ emission along the parsec-scale jet, which allows us to infer kinematic and dynamic properties of the NIR parsec-scale jet. The mass-loss rate inferred for the NIR jet is $\dot{M}_\mathrm{ejec}\sim10^{-4}\mathrm{\,M_\odot\,yr^{-1}}$ and the thrust is $\dot{P}\sim10^{-2}\mathrm{\,M_\odot\,yr^{-1}\,km\,s^{-1}}$ , which is roughly constant for the formation history of the young star. A tentative estimate of the ionisation fraction is derived for the massive jet by comparing the radio and NIR mass-loss rates. An ionisation fraction $\lesssim8\%$ is obtained, which means that the bulk of the ejecta is traced by the NIR jet and that the radio jet only delineates a small portion of it.}
        {}
    

   \maketitle
%


\section{Introduction}

In recent years, significant progress has been made in understanding the formation of high-mass young stellar objects (HMYSOs) (i.e. $M_*\geq8\mathrm{\,M_\odot}$, $L_\mathrm{bol}\geq5\times10^3\mathrm{\,L_\odot}$). The latest observational and theoretical studies present evidence that tips the balance of favour in a key debate over how high-mass stars form. It now seems likely that HMYSOs are born in the same way as their low-mass counterparts, via disc accretion, rather than through coalescence of lower mass stars. The growing observational evidence to support this view includes i) discovery of dusty discs around HMYSOs via near-infrared (NIR) interferometric image reconstruction \citep{kraus2010,kraus2017} and mid-infrared (MIR) interferometry \citep{boley2016}, ii) observations of HMYSO discs in Keplerian rotation through modelling of CO band-head emission \citep{ilee2013}, iii) direct detection of HMYSO discs in molecular tracers \citep{beltran2016}, iv) detection of several parsec-scale collimated infrared jets \citep{caratti2008,caratti2015,varricatt2010,cesaroni2013}, and v) discovery of the first disc-mediated accretion burst in a $20\mathrm{\,M_\odot}$ YSO \citep{caratti2017nature}. If, as now appears likely, HMYSOs form through the accretion-ejection process, then studies of the jet and outflow have the potential to provide valuable insights into their accretion processes as well as how these processes scale with the mass of the young stars. 

As HMYSOs are deeply embedded in their parental cloud, jet and outflow observations have traditionally relied on tracers at long wavelengths to minimise the effects of extinction. For example, HMYSO molecular outflows have been well studied in the molecular tracers of CO and SiO, observed in the sub-millimeter and millimeter regime \citep[see e.g.][]{beuther2002,maud2015}. However, these emission lines are generally considered tracers of the secondary outflow, that is, ambient material swept-up by the faster primary jet. Meanwhile, the primary jet has been well studied in radio emission \citep{guzman2012,masque2015,rosero2016,purser2016}. However, as radio emission traces only ionised gas, it may not trace the bulk of the ejecta since it depends on the degree of ionisation of the jet. Success in observing HMYSO primary jets at shorter wavelengths has been achieved, despite extinction, by moving to the NIR regime.
The primary jet of HMYSOs has also been seen to emit in several NIR atomic (\ion{H}{i}, [\ion{Fe}{ii}]) and molecular lines (H$_2$) \citep{davies2010,stecklum2012,cooper2013,caratti2015,caratti2016}, which trace the warm/hot gas of these shocked jets ($T>2000$\,K). The instantaneous efficiency at which mass is accumulated is measured by $\dot{M}_\mathrm{ejec}/\dot{M}_\mathrm{acc}$. Conceivably, this ratio could vary not only with evolutionary phase, but also with the mass of the central object. In turn, this could give us clues as to the underlying ejection mechanism, because massive jets could be a scaled-up version of their low-mass counterparts.

A limited number of kinematic studies of HMYSO primary jets exist in the NIR, the majority focusing on a single object \citep{davis2004,gredel2006,caratti2008,caratti2015,caratti2016}. Therefore, improving jet statistics is crucial to facilitate conclusions on disc-mediated accretion in HMYSOs, because the discs and stars themselves suffer high visual extinction, thereby hindering access to direct observations.
   
In this paper we report our findings on IRAS\,13481-6124 (G310.0135+00.3892), an HMYSO located at $3.1\pm1.1$\,kpc \citep{lumsden2013}. This object has a mass of $\sim20\mathrm{\,M_\odot}$ ($L_\mathrm{bol}=5.7\times10^4\mathrm{\,L_\odot}$), a spectral type of O9, and an age of $\sim6\times10^{4}$ years \citep{grave2009}. Detection of an accretion disc (inclination $\sim45\degr$, position angle $\sim 120 \degr$) was achieved via IR interferometry \citep{kraus2010,boley2016}. The central source drives a parsec-scale collimated bipolar jet with an extension of $\sim7$\,pc \citep{stecklum2012, caratti2015}. The parsec-scale jet has a precession angle of $\sim8\degr$ \citep[][see also Fig. \ref{fig:data_reduction}]{caratti2015}. By means of NIR interferometry on the Br$\gamma$ line, \citet{caratti2016} detected ejecta very close to the source ($\sim$few au) that extended for a few tens of au, suggesting that Br$\gamma$ is tracing the inner jet and that the protostar is still accreting material. This system has also been studied at radio wavelengths, and its radio jet dynamic properties have been derived~\citep{purser2016}. A mass-loss rate of $\sim1.8\times10^{-5}\mathrm{\,M_\odot\,yr^{-1}}$ and momentum rate of $\sim1-2\times10^{-2}\mathrm{\,M_\odot\,yr^{-1}\,km\,s^{-1}}$ were found \citep[assuming $v_\mathrm{jet}=500\mathrm{\,km\,s^{-1}}$ and $x_e=0.2$,][]{purser2016}, consistent with typical values determined for a sample of HMYSOs. Here, we present the first detailed study of the NIR jet of IRAS\,13481-6124. We examine the jet on small and large spatial scales, we compare it to the radio jet, and we thus present the first report on the kinematic and dynamic properties of the parsec-scale jet. 

This paper is structured as follows: in Sect. \ref{sect:observations} we describe our observations and in Sect. \ref{sect:data_analysis} we describe the data analysis. In Sect. \ref{sect:results} we present the results on source and along the jet at different spatial scales, including the kinematic and dynamic properties. Sect. \ref{sect:lte_model} presents our local thermal equilibrium
(LTE) model. Sect. \ref{sect:discussion} discusses the results we obtained. Finally, in Sect. \ref{sect:conclusions} our conclusions are presented.



\section{Observations}\label{sect:observations}

Observations of IRAS\,13481-6124 were obtained using three ESO-VLT
(Very Large Telescope) instruments: the spectrograph for integral field observations in the near-infrared \citep[SINFONI,][]{sinfoni2003}; the cryogenic high-resolution pre-dispersed infrared echelle spectrograph \citep[CRIRES,][]{crires2004}, and the infrared spectrometer and array camera \citep[ISAAC,][]{isaac1998}. Details of the observations can be found in Table \ref{tab:observations}. 

Each instrument was used to study different regions of the IRAS\,13481-6124 system at different scales: SINFONI was used to analyse the source and its immediate environment, and CRIRES and ISAAC were used to study the jet; the former focusing on the au-scale jet using the technique of spectro-astrometry and the latter focusing at the parsec-scale jet investigating its dynamic and kinematic properties. 

\subsection{SINFONI data}

VLT/SINFONI integral field unit (IFU) observations were obtained on 2011 April 27 (Program ID 087.C-0951(B)) in $K$ band. The field of view of $8''\times8''$ was centred on the source, with a position angle (PA) east of north (E of N) of zero degrees. Spatial sampling was $125\times250$ mas/pixel, with the smaller sampling in the northern direction. Total integration time was 3\,s. Spectral resolution was $\mathcal{R}\sim4\,000$ ($75\,\mathrm{km\,s^{-1}}$), and spatial resolution achieved using AO+NGS was $0.1''$. The natural guide star used for the AO system was 2MASS J13513620-6138563 ($V=12.3,R=12.6$ mag and separation of $16''$ from the target).

Additional SINFONI data were obtained on 2012 February 21 and 22 (Program ID  088.C-0575(C)) in $H$ and $H+K$ bands. The field of view of $3''\times3''$ was centred on the source, with a PA of $-31\degr$ in both bands. Spatial sampling was $50\times100$ and $125\times250$ mas/pixel for the $H$ and $H+K$ bands, respectively, with the smaller sampling in the jet direction. Total exposure time was 8 and 16\,s for the $H$ band (both were coadded to increase the signal-to-noise ratio,  S/N) and $2$\,s for $H+K$ band. The spectral resolution was $\mathcal{R}\sim3\,000$ and 1\,500 (100 and 200$\,\mathrm{km\,s^{-1}}$), respectively. Spatial resolution of $0.5''$ (AO+NGS) was achieved for the $H$ band and $1.0''$ (seeing-limited) for the $H+K$ band. 

All data were reduced in the standard way, using dedicated instrument software, {\scriptsize GASGANO}, and standard IRAF routines for SINFONI. A wavelength accuracy of $0.31$\AA\,(or $\sim4.2-5.6\mathrm{\,km\,s^{-1}}$ for the $H$ and $K$ bands, respectively) was achieved. Flux calibration was performed using the photometric standard star Hip072690.

\subsection{CRIRES data}

VLT/CRIRES high spectral resolution observations were carried out on 2013 March 10, 15, and 16 in the $L$, $K$, and $J$ bands, respectively (Program ID  090.C-0371(C)). A long slit of $0.4''$ $\times$ $40''$ was centred on Br$\alpha$ ($\lambda_{\mathrm{vac}}=4.05226\,\mu$m), Br$\gamma$ ($\lambda_{\mathrm{vac}}=2.16612\,\mu$m), and Pa$\beta$ ($\lambda_{\mathrm{vac}}=1.28216\,\mu$m). This set-up achieved a spectral resolution of $\mathcal{R}\sim50\,000$ ($6\,\mathrm{km\,s^{-1}}$). Spatial sampling was 86 mas/pixel. The slit was placed first along the jet ($204^{\circ}$), and then perpendicular to the jet ($114^{\circ}$). The slit was also placed at two anti-parallel PA values ($24^{\circ}$ and $294^{\circ}$). This observing strategy allows for the identification and removal of artefacts before applying the technique of spectro-astrometry (see Sect. \ref{sect:data_analysis}). Use of AO+NGS achieved spatial resolutions of $0.35''$, $0.3''$, and $0.2''$ for the Pa$\beta$, Br$\gamma$, and Br$\alpha$ lines, respectively. The NGS used in CRIRES observations was the same as in SINFONI observations.

The data were reduced in the standard way using the {\scriptsize GASGANO} package, and following the CRIRES data reduction cookbook. A wavelength accuracy of 0.22, 0.25, and 0.61\,\AA\, (or $\sim$ 5.1, 3.5, and 4.5$\mathrm{\,km\,s^{-1}}$ for the $J$, $K$, and $L$ bands, respectively) was achieved. The telluric standard star used was HR5071 to remove atmospheric features, for the Pa$\beta$ and Br$\gamma$ spectra. In the case of the Br$\alpha$ spectrum, due to the poor quality of the standard star in the $L$ band, telluric features were modelled and removed using templates of the Earth's telluric features. CRIRES spectra were not flux calibrated.

\subsection{ISAAC data}

\begin{figure}[ht]
        \centering
        \includegraphics[width=0.5\textwidth]{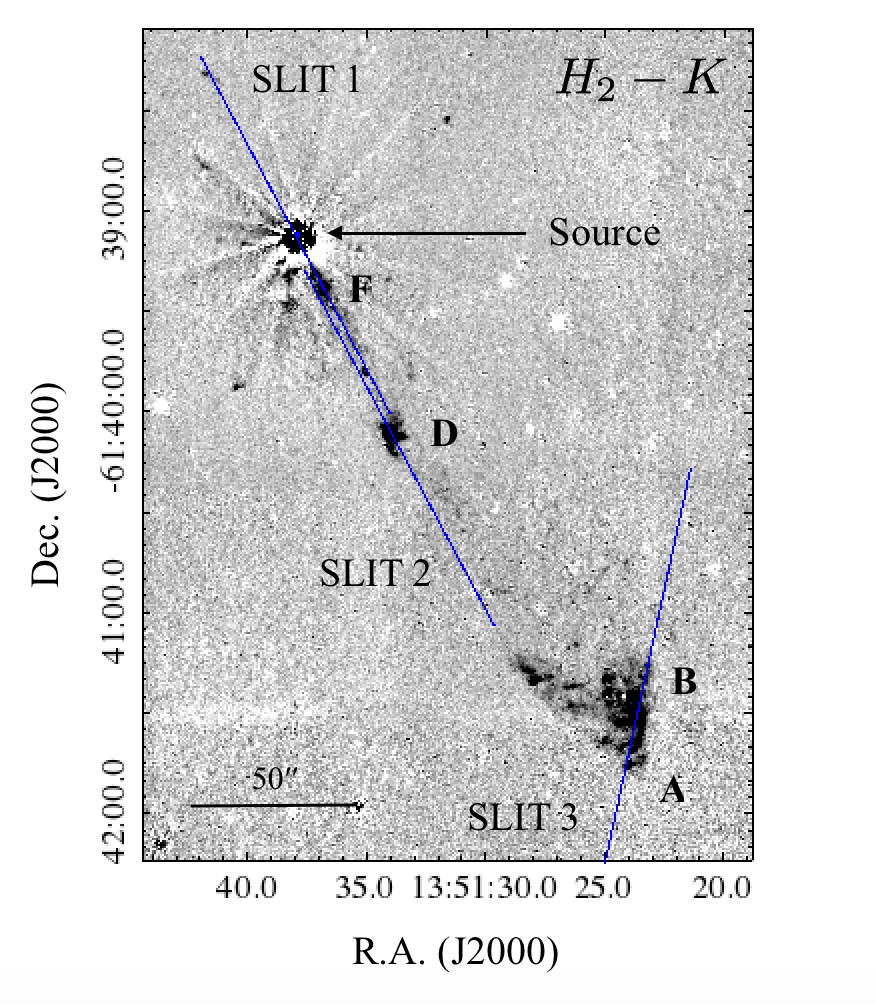}
        \caption{SOFI\protect\footnotemark\, H$_2$ continuum-subtracted image of IRAS\,13481-6124 showing the blue-shifted jet knots, bow shock, and driving source. Knot positions are labelled A, B, D, and F, following the notation of \citet{caratti2015}. ISAAC slit positions (blue) are labelled SLIT 1, 2, and 3 for convenience,  and correspond to observations on the source, jet, and bow shock, respectively.} 
        \label{fig:data_reduction}
\end{figure}

High-resolution long-slit spectra were obtained with VLT/ISAAC on 2011 April 18 in $H$ and $K$ bands (Program ID 087.C0951(A)). High-resolution spectra around the [\ion{Fe}{ii}] ($\lambda_{\mathrm{vac}}=1.6642\,\mu$m), (from 1.4 to 1.82 $\mu$m), and around the $1-0$\,S(1) H$_2$ ($\lambda_{\mathrm{vac}}=2.12183\,\mu$m), and Br$\gamma$ ($\lambda_{\mathrm{vac}}=2.16612\,\mu$m) lines were obtained. The $0.3'' \times 120''$ long slit was positioned at three orientations: on-source (PA $=-152.5\degr$; slit 1, see Fig. \ref{fig:data_reduction}), along the jet (PA $=27.5\degr$; slit 2, see Fig. \ref{fig:data_reduction}), and on the terminal bow shock (PA $=167.5\degr$; slit 3, see Fig. \ref{fig:data_reduction}). Spatial sampling was 146 mas/pixel. Total integration time was 180 and 150\,s for $H$ and $K$ band, respectively. Spectral resolution was $\mathcal{R}\sim10\,000$ and 8\,900 ($30-35\mathrm{\,km\,s^{-1}}$) for the $H$ and $K$ bands, respectively. Spatial resolution was seeing-limited ($0.8-1.0''$). The standard $ABBA$ nodding strategy was used.

Data were reduced in the standard way using IRAF. Wavelength calibration relied on the OH atmospheric lines in each frame. The spatial distortion and curvature caused by the long slit were corrected using the calibration file {\scriptsize STARTRACE}. A wavelength accuracy of $0.061$\AA\, (or $\sim1.1 \mathrm{\,and\,} 0.8\mathrm{\,km\,s^{-1}}$ for the $H$ and $K$ bands, respectively) was achieved. Atmospheric telluric lines were corrected for by observing the telluric standard star Hip058630. Additionally, the on-source spectra were flux calibrated using the photometric standard star Hip058630.

\footnotetext{Son OF Isaac is a low-resolution spectrograph and imaging camera on the ESO New Technology Telescope (NTT).}

\begin{table*}
\caption{Summary of the observations carried out with VLT/SINFONI, ISAAC, and CRIRES.}             
\label{tab:observations}      
\centering          
\begin{tabular}{c c c c c c c c}     
\hline\hline       
\noalign{\smallskip}
Date of Obs. & Telescope/ & Wavelength & DIT & PA & Resolution & Target\\ 
(dd.mm.yyyy) & Instrument & ($\mu\mathrm{m}$) & (s) & ($\degr$) & $\mathcal{R}$ &  \\
\noalign{\smallskip}
\hline              
\noalign{\smallskip}
   27.04.2011 & VLT/SINFONI & 2.2 & 3 & 0 & 4000 & Source\\
   21.02.2012 & VLT/SINFONI & 1.65-2.2 & 2 & -31 & 1500 & Source\\
   22.02.2012 & VLT/SINFONI & 1.65 & 8 & -31 & 3000 & Source\\
   22.02.2012 & VLT/SINFONI & 1.65 & 16 & -31 & 3000 & Source\\
   18.04.2011 & VLT/ISAAC & 1.65 & 180 & -152.5 & 10000 & Knot D, F \\  
   18.04.2011 & VLT/ISAAC & 2.2 & 150 & 27.5 & 8900 & Knot D, F \\
   18.04.2011 & VLT/ISAAC & 2.2 & 150 & 167.5 & 8900 & Knot A, B\\
   18.04.2011 & VLT/ISAAC & 2.2 & 10 & 27.5 & 8900 & Source\\
  10.03.2013 & VLT/CRIRES & 4.05 & 5 & 24 & 50000 & On source\\
  10.03.2013 & VLT/CRIRES & 4.05 & 5 & 114 & 50000 & On source\\
  10.03.2013 & VLT/CRIRES & 4.05 & 5 & 204 & 50000 & On source\\
  10.03.2013 & VLT/CRIRES & 4.05 & 5 & 294 & 50000 & On source\\
  15.03.2013 & VLT/CRIRES & 2.16 & 20 & 24 & 50000 & On source\\
  15.03.2013 & VLT/CRIRES & 2.16 & 20 & 114 & 50000 & On source\\
  15.03.2013 & VLT/CRIRES & 2.16 & 20 & 204 & 50000 & On source\\
  15.03.2013 & VLT/CRIRES & 2.16 & 20 & 294 & 50000 & On source\\
  16.03.2013 & VLT/CRIRES & 1.28 & 90 & 24 & 50000 & On source\\
  16.03.2013 & VLT/CRIRES & 1.28 & 90 & 114 & 50000 & On source\\
  16.03.2013 & VLT/CRIRES & 1.28 & 90 & 204 & 50000 & On source\\
  16.03.2013 & VLT/CRIRES & 1.28 & 90 & 294 & 50000 & On source\\
\noalign{\smallskip}
\hline                  
\end{tabular}
\end{table*}


\section{Data analysis}\label{sect:data_analysis}

To generate the SINFONI spectra, a region of $0.5''\times0.5''$ (i.e., $1550\times1550$\,au at a distance of 3.1\,kpc) was extracted from the data cubes. ISAAC and CRIRES spectra were extracted using specific IRAF tasks for long slit. The wavelength calibration was carried out via two different methods: for ISAAC and CRIRES spectra, atmospheric telluric lines were used; and for SINFONI spectra, calibration relied on arc lamp observations.

Radial velocities of the [\ion{Fe}{ii}], Br$\gamma$, and H$_2$ lines were measured from the ISAAC spectra. These lines are used to compute kinematic and dynamic properties shown in Sect. \ref{sect:results}. The emission lines were fitted by a Gaussian profile with a typical error of $5-6\mathrm{\,km\,s^{-1}}$, given the high S/N ($\sim100$ for the brightest lines). All velocities in ISAAC and CRIRES spectra are with respect to the local standard of rest (LSR), and were corrected for the velocity of the parental cloud \citep[$-37.9\mathrm{\,km\,s^{-1}}$,][]{lumsden2013}.

\subsection{Spectro-astrometry}\label{sect:data_analysis_SA}

Given the high spectral resolution of CRIRES and the excellent S/N of our observations, we are in a position to apply the technique of spectro-astrometry in order to retrieve spatial information on scales below the effective resolution of the observations \citep{bailey1998,takami2001,takami2003,whelan2005}. The technique involves measuring the spatial offset of the emission centroid with respect to that of the continuum as a function of wavelength. The accuracy, $\sigma$, of the centroid position, well below the effective spatial resolution, is given by \citet{whelan2008}, nonetheless, the error given by this formula is likely underestimated:

\begin{equation}
\sigma=\frac{\mathrm{seeing}}{2.3548\,\sqrt{\mathrm{N_p}}},
\end{equation}

\noindent where the seeing is measured in the spatial direction and is determined by the FWHM of the observations (i.e. the FWHM of the continuum). $\mathrm{N_p}$ is the number of photons at the peak of the line. Given the spectral resolution and S/N in our CRIRES observations, this equation provides an accuracy of up to $\sim0.65$\, milliarcsecond (mas) in our case, which implies that we can probe our target on au scales. Although the technique is powerful, it is prone to contamination by artefacts \citep{bailey1998,takami2001,whelan2008}. For example, the spectro-astrometric signal is affected by a misalignment of the detector with respect to the spatial direction, and therefore a$\text{ second-order}^\mathrm{}$ polynomial was fitted to correct for this distortion. To remove artefacts, the parallel (p) and the antiparallel (ap) slit measurements were subtracted from each other for each pixel $(x)$ \citep{brannigan2006}, as shown in Eq. \ref{eq:rem_art},

\begin{equation}
Y(x) = (Y_\mathrm{p}(x)-Y_\mathrm{ap}(x))/2
\label{eq:rem_art}
.\end{equation}

A fitting of the continuum was not possible due to the very broad nature of the lines, but instead we were able to estimate the continuum intensity level by examining the small portion of the continuum to the left of the line. The measured line displacement was corrected by multiplying by the following weight in order to derive the offset of the line emitting region from the measured combined line and continuum emission offset \citep[see e.g.][]{takami2001}:

\begin{equation}
W(\lambda)=\frac{I_{\lambda(\mathrm{line})}+I_{\lambda(\mathrm{cont})}}{I_{\lambda(\mathrm{line})}},
\label{eq:weights}
\end{equation}

\noindent where $I_{\lambda\mathrm{(line)}}$ is the continuum-subtracted intensity in the line.

Finally, it is also important to correct for photospheric features. However, IRAS\,13481-6124 is probably an O9 star \citep[see][]{grave2009}, and the \ion{H}{i} photospheric contribution should be negligible. Nevertheless, we examined our data for such features (see Sect. \ref{sect:results_source}).


\section{Results}\label{sect:results}

We have divided the results into three subsections, corresponding to the three main regions of our target: 
the source and its immediate circumstellar environment; the base of the jet, that is, the au-scale jet; and the parsec-scale jet.

\subsection{Source and its immediate environment}\label{sect:results_source}

\begin{figure*}[ht]
        \centering
        \includegraphics[width=1.0\textwidth]{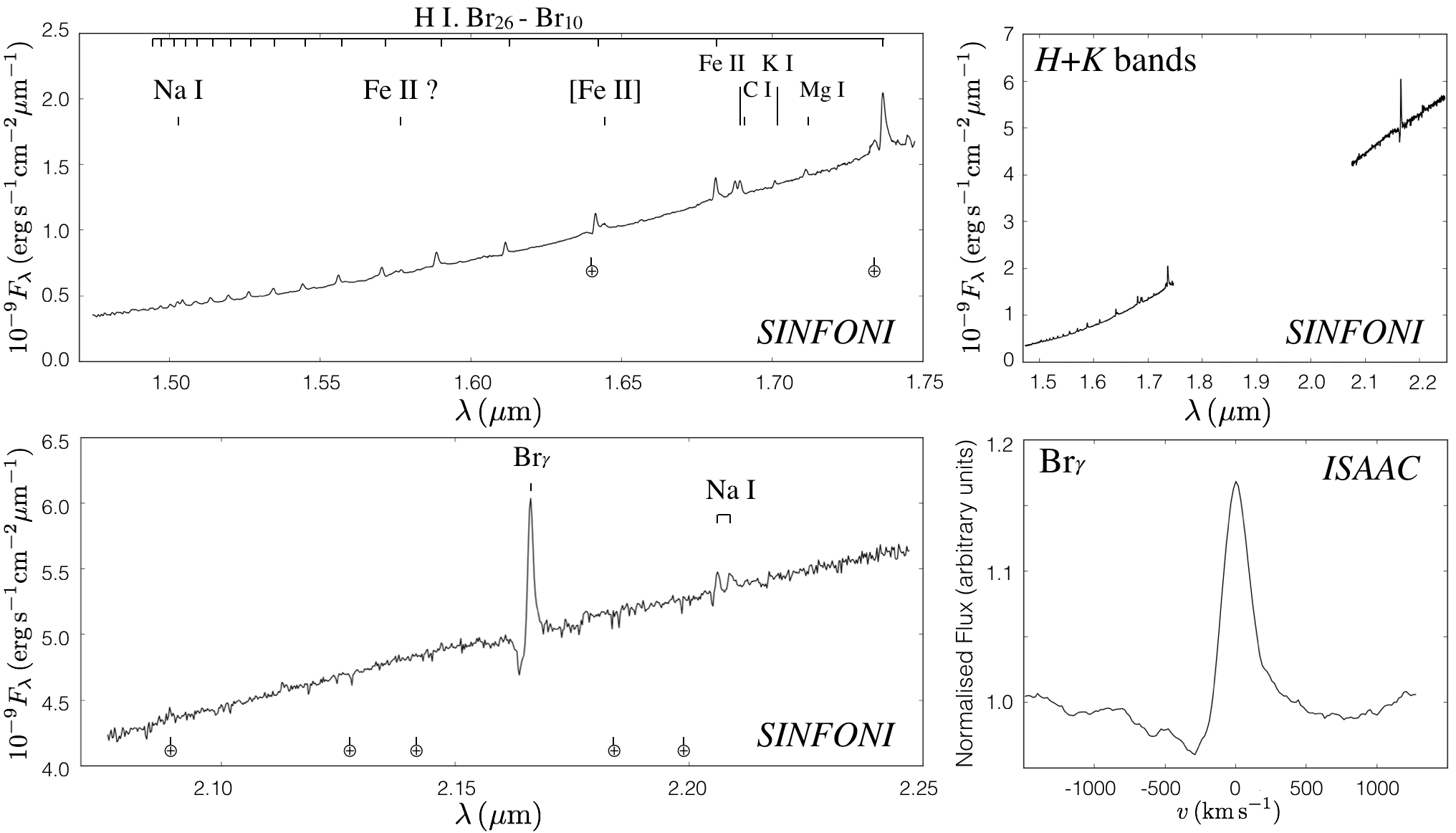}
        \caption{\textit{Top left:} $H$ -band SINFONI spectrum on source. Detected lines and telluric features ($\oplus$) are labelled. \textit{Bottom left:} $K$ -band SINFONI spectrum on source. \textit{Top right:} SED from the $H$ and $K$ spectra. \textit{Bottom right:} Zoom-in of the Br$\gamma$ line of the ISAAC spectrum. Velocities are with respect to the local standard of rest (LSR).}
        \label{fig:spectra}
\end{figure*}

Several emission lines on source and in the immediate environment are detected in both SINFONI and ISAAC data (see Table \ref{tab:lines} and Fig. \ref{fig:spectra} upper and lower left panels). These lines mainly trace circumstellar features. Although the observed emission lines are not spatially resolved, they have different excitation energies and therefore trace a variety of processes at distinct spatial scales.
In the $H$ band, we detect the Brackett series lines (from Br26\,1.4941 $\mu$m to Br10\,1.7376 $\mu$m), which are mainly associated with accretion and ejection processes \citep{muzerolle1998,garcia-lopez2006,caratti2015}. In addition, permitted and forbidden atomic lines are also identified. In particular, \ion{Fe}{ii} (1.6811 $\mu$m), \ion{C}{i} (1.6894 $\mu$m), and \ion{Mg}{i} (1.7113 $\mu$m) are believed to originate from chromospheric activity \citep{hartmann1992a,hartmann1992b,kelly194}. However, in the early stage of HMYSOs, there is no clear evidence of the existence of chromospheres. Hence, these lines may trace fluorescent emission. Indeed, OB stars emit sufficient UV photons to pump those lines by fluorescence. The [\ion{Fe}{ii}] line traces the base of the jet very close to the central engine \citep{nisini2002,caratti2006}. In the $K$ band, we detect the prominent Br$\gamma$ line at 2.1662 $\mu$m, as well as the \ion{Na}{i} doublet at 2.2062/2.2089 $\mu$m, which originates in the disc \citep{lorenzetti2011}. Detected lines and corresponding fluxes are listed in Table \ref{tab:lines}. Interestingly, no photospheric features were detected in any of our spectra from all three instruments. This is probably due to the high veiling. From our ISAAC spectra we estimate $r_K\gtrsim 35$. Therefore, expected photospheric lines such as \ion{He}{i} lines at 1.571, 1.691, and 1.700 $\mu$m in the $H$ band \citep{blum1997}, or \ion{He}{i} lines at 2.058, 2.112, 2.113 $\mu$m, \ion{He}{ii} 2.185 $\mu$m, \ion{Ne}{iii} 2.115 $\mu$m, and the \ion{C}{iv} at 2.069, 2.078, 2.083 $\mu$m in the $K$ band \citep{bik2005b} cannot be seen. As a consequence, we are not able to confirm the spectral type of the central source. 

Figure \ref{fig:spectra} (bottom right) zooms in on the Br$\gamma$ line of the $K$ band ISAAC spectrum. Notably, the Br$\gamma$ line of IRAS\,13481-6124 displays a P\,Cygni profile. Such an asymmetric line profile is indicative of an absorbing blue-shifted outflow with a measurable bulk velocity. We measure for the centroid of the blue absorption of the P-Cygni profile for the Br$\gamma$ $v_\mathrm{rad}\sim-290\pm40\mathrm{\,km\,s^{-1}}$ (which corresponds to a total velocity of $v_\mathrm{tot}\sim410\mathrm{km\,s^{-1}}$, considering an inclination angle of $45\degr$), which is tracing the bulk of the material of the wind close to the star. This velocity is likely very close to the terminal velocity of the jet. Velocities measured at large scales (with H$_2$ and [\ion{Fe}{ii}] emission) are similar to the velocity found close to the star. We also measure a radial velocity peak of $\sim10\pm5\mathrm{\,km\,s^{-1}}$ and a full width at zero intensity (FWZI) of $\sim900\mathrm{\,km\,s^{-1}}$.

Figure \ref{fig:spectra} (upper right panel) shows the spectral energy distribution (SED) of IRAS\,13481-6124 in the $H$ and $K$ bands. An increase in continuum flux towards the longer wavelengths is observed. This IR-excess is due to the combined effect of dust extinction and dust emission towards the source. 

\begin{table}
\caption{Observed emission lines on IRAS13481-6124.}             
\label{tab:lines}      
\centering          
\begin{tabular}{c c c c }    
\hline\hline       
\noalign{\smallskip}
Species & Transition & $\lambda_\mathrm{vac}$ & Flux  \\ 
  & &($\mu\mathrm{m}$) &  ($10^{-14}\mathrm{\,erg\,cm^{-2}\,s^{-1}}$) \\
\noalign{\smallskip}
\hline              
\noalign{\smallskip}
   \ion{H}{i} & Br$_{26}$ & 1.4941 & $2.0\pm0.2$ \\
   \ion{H}{i} & Br$_{25}$ & 1.4971 & $2.4\pm0.3$ \\
   \ion{H}{i} & Br$_{24}$ & 1.5004 & $3.0\pm0.3$ \\
   \ion{Na}{i} & ${}^2D_{5/2} - {}^2F_{7/2} $ & 1.5026 & $4.0\pm0.1$ \\
   \ion{H}{i} & Br$_{23}$ & 1.5043 & $4.3\pm0.1$ \\
   \ion{H}{i} & Br$_{22}$ & 1.5086 & $4.4\pm0.4$ \\
   \ion{H}{i} & Br$_{21}$ & 1.5137 & $5.8\pm0.3$ \\
   \ion{H}{i} & Br$_{20}$ & 1.5196 & $6.6\pm0.3$ \\
   \ion{H}{i} & Br$_{19}$ & 1.5264 & $6.9\pm0.4$ \\
   \ion{H}{i} & Br$_{18}$ & 1.5346 & $7.8\pm0.7$ \\
   \ion{H}{i} & Br$_{17}$ & 1.5443 & $7.2\pm1.0$ \\
   \ion{H}{i} & Br$_{16}$ & 1.5560 & $8.8\pm1.0$ \\
   \ion{H}{i} & Br$_{15}$ & 1.5705 & $10.0\pm0.9$ \\
   \ion{Fe}{ii} ? & ${}^2F_{7/2}-{}^2D_{5/2}$ & 1.5768 & $3.8\pm1.2$ \\
   \ion{H}{i} & Br$_{14}$ & 1.5884 & $14.6\pm1.1$ \\
   \ion{H}{i} & Br$_{13}$ & 1.6113 & $13.4\pm1.7$ \\
   \ion{H}{i} & Br$_{12}$ & 1.6411 & $16.0\pm2.4$ \\
   $[$\ion{Fe}{ii}$]$ & $\mathrm{a^4D_{7/2}-a^4F_{9/2}}$ & 1.6440 & $9.1\pm2.6$ \\
   \ion{H}{i} & Br$_{11}$ & 1.6811 & $21.9\pm1.7$ \\
   \ion{Fe}{ii} & z4Fo-c4F & 1.6877 & $13.3\pm1.6$ \\
   \ion{C}{i} & 1D-1F0 2 - 3 & 1.6894 & $12.9\pm1.6$ \\
   \ion{K}{i} & ${}^2S_{1/2,1/2}-{}^2P_{1/2,3-2}$ & 1.7010 & $3.7\pm0.7$ \\
   \ion{Mg}{i} & ${}^1P_1-{}^1S_0$ & 1.7113 & $6.6\pm1.0$ \\
   \ion{H}{i} & Br$_{10}$ & 1.7366 & $41.8\pm4.6$ \\
   \ion{H}{i} & Br$\gamma$ & 2.1661 & $192\pm5.5$ \\
   \ion{Na}{i} & ${}^2S_{3/2}-{}^2Po_{1/2}$ & 2.2062 & $17.1\pm3.2$ \\
   \ion{Na}{i} & ${}^2S_{1/2}-{}^2Po_{1/2}$ & 2.2089 & $24.5\pm4.3$ \\

\noalign{\smallskip}
\hline                  
\end{tabular}
\end{table}

\subsection{Jet base: \ion{H}{i} emission lines and their spatial displacements}

\begin{figure}[ht]
\centering
\includegraphics[width=0.5\textwidth]{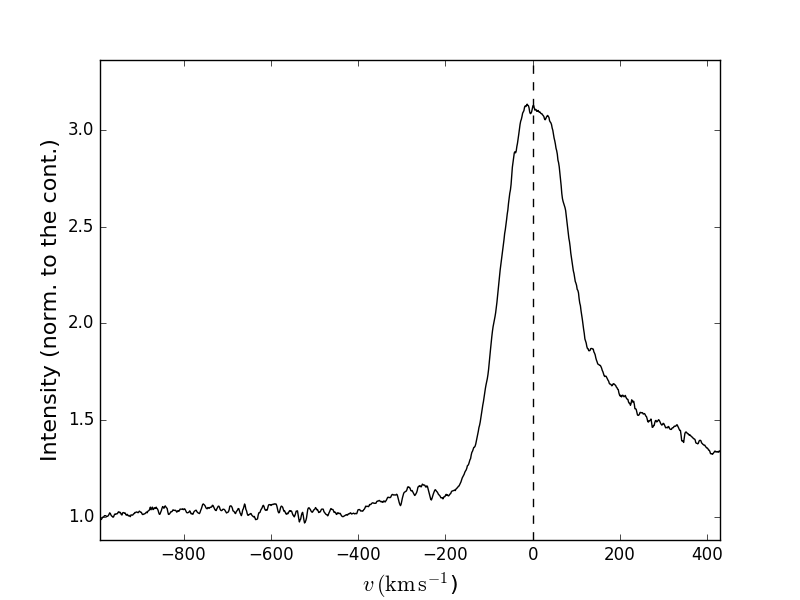}
\caption{CRIRES spectrum of the Pa$\beta$ line. Intensity is normalised to the continuum. Velocity is with respect to the LSR.}
\label{fig:PaB_line}%
    \end{figure}

CRIRES spectro-astrometry on the Pa$\beta$, Br$\gamma$, and Br$\alpha$ lines was performed to retrieve photocentre shifts of the line with respect to the continuum on mas scales (see Sect. \ref{sect:data_analysis}). Figure \ref{fig:PaB_line} shows the spectrum of the Pa$\beta$ line. Part of the line is missing due to the wavelength coverage and the gaps between the chips. Nonetheless, the shape of the line is clearly asymmetric, that is, the blue-shifted wing is absent. This effect is produced by self-absorption along the flow caused by the strong wind. The FWZI of the Pa$\beta$ line is $\sim900\mathrm{\,km\,s^{-1}}$, consistent with what we find with ISAAC for the Br$\gamma$ line. This would then suggest not just a similar wind origin, but also that it is produced in the same region. Unfortunately, no significant spectro-astrometric signal was detected on the Pa$\beta$ line due to the low S/N $(\sim7)$ of the Pa$\beta$ line. The upper limit on the spectro-astrometric signal is defined by the detection limit, which is $\sim23$ mas.

\begin{figure*}[ht]
        \centering
    \includegraphics[width=0.49\textwidth]{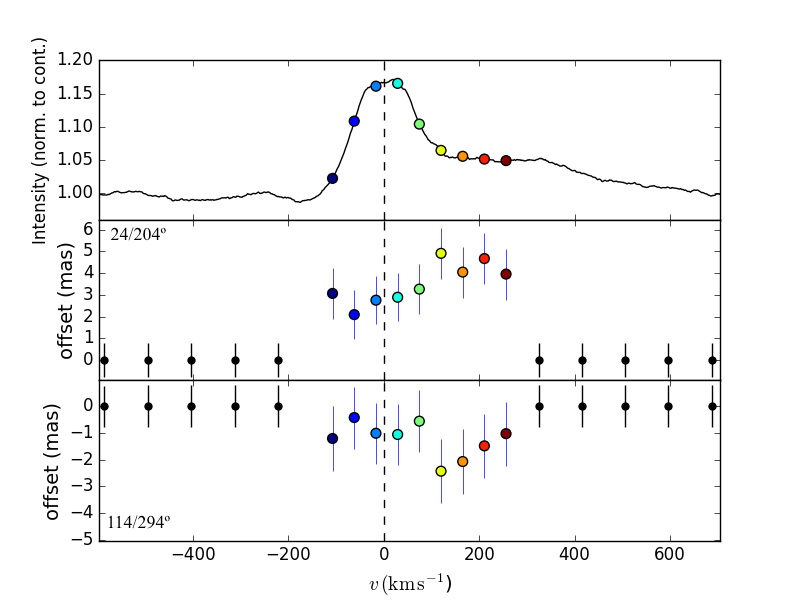}
        \includegraphics[width=0.49\textwidth]{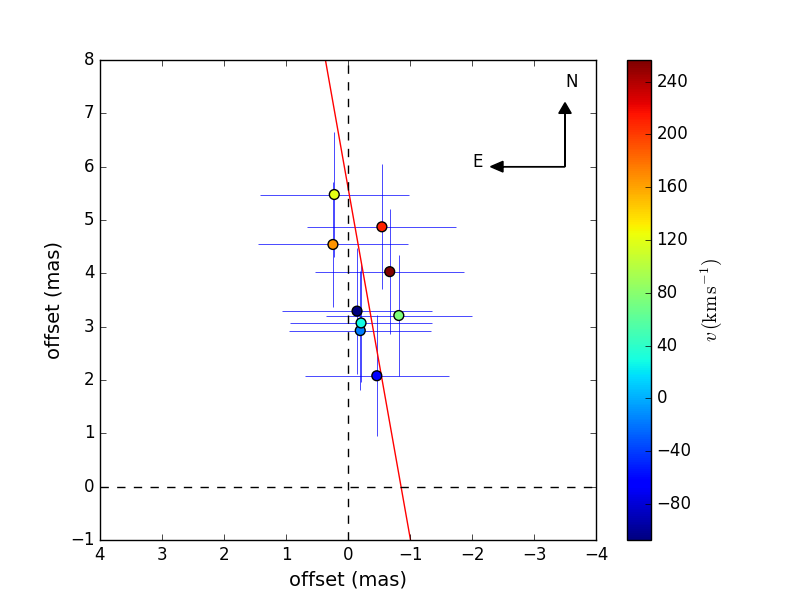}
        \caption{\textit{Left panel:} CRIRES Br$\gamma$ spectrum (top) along with the spectro-astrometric signal (in mas) detected in the jet parallel (middle) and perpendicular slit (bottom). The spectro-astrometric signal is continuum-corrected. \textit{Right panel:} Spectro-astrometric signal on the plane of the sky (north is up and east is to the left). The red line is the linear fit to the points giving a PA for the line-emitting region of $190\pm15\degr$. Error bars shown in both plots are $3\sigma$. Velocities are colour coded and are the same in both panels.}
        \label{fig:BrG_disp}
\end{figure*}

\begin{figure*}[ht]
        \centering
    \includegraphics[width=0.49\textwidth]{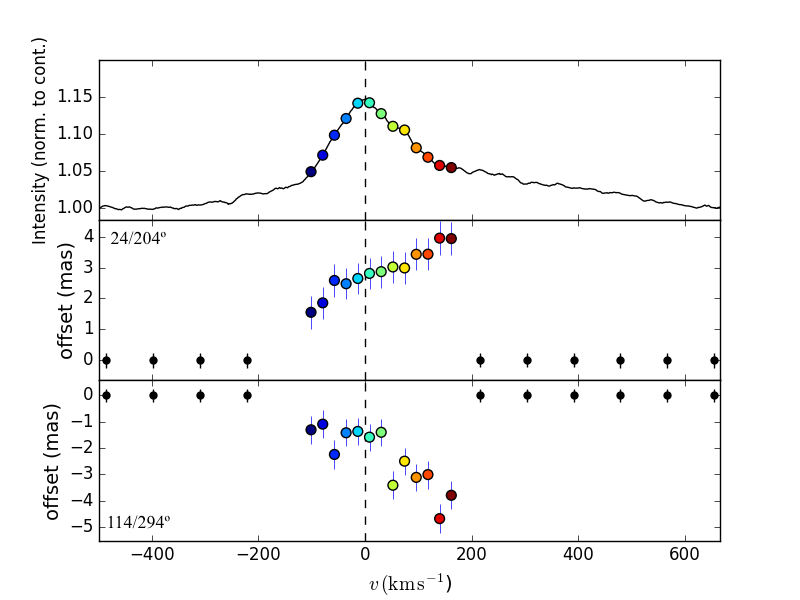}
        \includegraphics[width=0.49\textwidth]{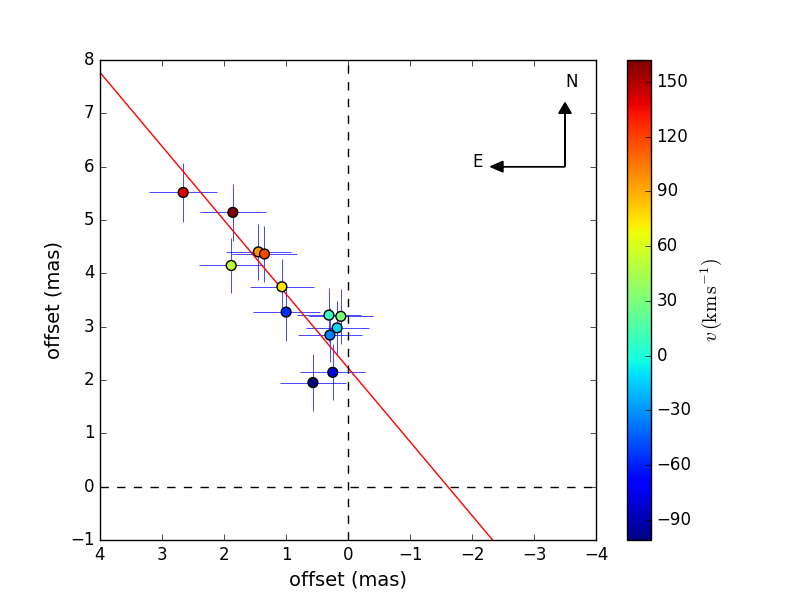}
        \caption{\textit{Left panel:} CRIRES Br$\alpha$ spectrum (top) along with the spectro-astrometric signal (in mas) detected in the jet parallel (middle) and perpendicular slit (bottom). The spectro-astrometric signal is continuum-corrected. \textit{Right panel:} Spectro-astrometric signal on the plane of the sky (north is up and east is to the left). The red line is the linear fit to the points giving a PA for the line-emitting region of $216\pm5\degr$. Error bars shown in both plots are $3\sigma$. Velocities are colour coded and are the same in both panels.}
        \label{fig:BrA_disp}
\end{figure*}

By contrast, the Br$\gamma$ and Br$\alpha$ lines (S/N ratio $\sim$ 62 and $\sim$130, respectively) show a significant spectro-astrometric signal. To improve the S/N of these lines, emission was binned spectrally: 32 channels (or pixel columns) were binned in the case of Br$\gamma$, and 16 for Br$\alpha$, corresponding to a velocity resolution of $55\mathrm{\,km\,s^{-1}}$ and $38\mathrm{\,km\,s^{-1}}$, respectively. Centroid positions were measured for points above $3\sigma$ of the continuum intensity, namely in the velocity ranges $\sim-110$ to $250\mathrm{\,km\,s^{-1}}$ for the Br$\gamma$ line and $\sim-100$ to $160\mathrm{\,km\,s^{-1}}$ for the Br$\alpha$ line (Figs.~\ref{fig:BrG_disp} and \ref{fig:BrA_disp} upper left panels). 
 
The Br$\gamma$ line offsets (continuum-compensated, as discussed in Sect. \ref{sect:data_analysis_SA}) extend up to $\sim4.5$\,mas ($\sim14$\,au) in the direction of the jet (i.e. parallel slit observations; Fig. \ref{fig:BrG_disp} middle left panel) and up to $\sim2.5$\,mas ($\sim8$\,au) in the direction orthogonal to the jet (i.e. perpendicular slit observations; Fig. \ref{fig:BrG_disp} bottom left panel). Figure \ref{fig:BrG_disp} right panel shows the centroid offset in the plane of the sky. The offset of the Br$\gamma$ line emitting region with respect to the continuum in the plane of the sky is $\sim3.5$\,mas ($\sim11$\,au), which is a lower limit because with spectro-astrometry one does not spatially resolve the emission. The red line is the linear fit to the points. From the slope of the fit ($m$) one can derive the PA of the jet ($\alpha_\mathrm{PA_{jet}}=\arctan(m)$). A value of $\alpha_\mathrm{PA_{jet}}\approx190\pm15\degr$ was found for the Br$\gamma$ line. This result is in good agreement with previously derived values for the IRAS\,13481-6124 jet \citep{kraus2010,stecklum2012, caratti2015}, confirming that the Br$\gamma$ is displaced along the jet axis (Fig. \ref{fig:BrG_disp} right panel).

Figure \ref{fig:BrA_disp} shows the spectrum (upper left panel) and, for the first time in a HMYSO, the centroid offsets of the Br$\alpha$ line (middle and lower left panel). The line profile is clearly asymmetric, as seen in the previous \ion{H}{i} lines. The FWZI of the three \ion{H}{i} lines investigated with CRIRES are identical and coincident with the one obtained with ISAAC for the Br$\gamma$ ($\sim900\mathrm{\,km\,s^{-1}}$).
As with the Br$\gamma$ line, the Br$\alpha$ emission extends along the jet direction, up to $\sim5$ mas ($\sim15$ au), which is again a lower limit. The blue-shifted wing of the Brackett lines extend towards to the south-west, and the red-shifted wing extend towards to the north-east. The magnitude of the centroid offset increases with increasing radial velocity; the blue-shifted wing offset reaches $\sim1.5$ mas at $\sim-100\mathrm{\,km\,s^{-1}}$, while the red shifted wing reaches $\sim5$ mas at $\sim160\mathrm{\,km\,s^{-1}}$ (Fig. \ref{fig:BrA_disp} right panel). These spectro-astrometric features reveal the presence of a bipolar jet close to the star \citep{takami2001}, as previously detected through interferometric observations of the Br$\gamma$ line \citep{caratti2016}. The red line indicates the linear fit to the Br$\alpha$ points. A PA of $\alpha_\mathrm{PA_{jet}}\approx216\pm5\degr$ is derived. Reassuringly, both Brackett lines yield a similar jet PA and agree with previous studies, which relied on different techniques \citep[][]{kraus2010,stecklum2012, caratti2015, caratti2016}. 

However, it is worth noting that our spectro-astrometric results show a shift towards the north north-east with respect to the position of the central source, namely position  (0,0) in Figures \ref{fig:BrG_disp} and \ref{fig:BrA_disp}. We would expect the red-shifted and blue-shifted jet offsets to straddle the continuum position. The simplest explanation is to consider that this effect is produced by a hypothetical binary companion in the surroundings of IRAS\,13481-6124. However, there is no indication of a close (mas-scale) companion from NIR interferometry data \citep{kraus2010,caratti2016}. In addition, there is no evidence of a companion in our SINFONI+AO observations, which achieved high spatial resolution (hundreds of mas). Finally, neither ISAAC nor CRIRES data show detection of a possible companion in the direction of the centroid offset. It would be detected as a second continuum in the CRIRES spectral images, which is not the case. Therefore, the spectro-astrometric offsets seem to have a different origin, such as an asymmetric FOV-dependent continuum distribution. We note that while the technique of spectro-astrometry cannot spatially resolve the emission, it does give information on the flux distribution. Conversely, interferometry resolves the jet at small (au) scales, but observes only a very small FOV around the star. As a consequence, it may partially resolve out extended, asymmetric continuum flux distribution. This explains why the AMBER Br$\gamma$ emission is observed to be centred on the source, while the CRIRES spectro-astrometry  reveals an offset. This offset can be explain by a contribution of large-scale circumstellar nebulosity. 

\subsection{Parsec-scale jet}

\begin{figure*}[ht]
\centering
\includegraphics[width=0.90\textwidth]{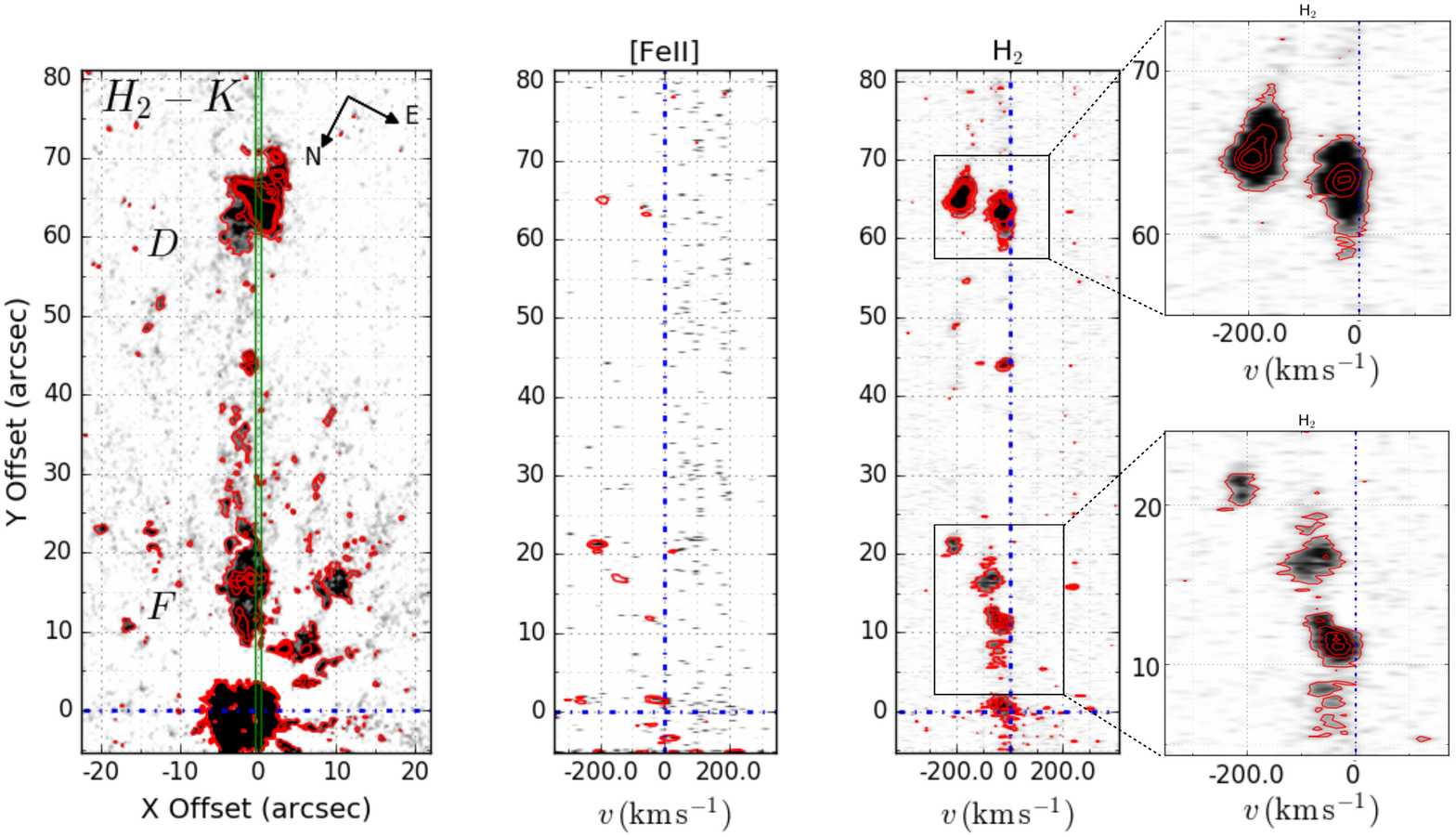}
\caption{PV diagram of slit 2. \textit{Left}: SOFI H$_2-K$ continuum subtracted image \citep{caratti2015}. Contours are from $3\sigma$ to $20\sigma$ in steps of $5\sigma$. \textit{Centre}: [\ion{Fe}{ii}] (1.644\,$\mu$m) line PV diagram tracing the atomic jet. Contours range from 3$\sigma$ to 5$\sigma$ in steps of 1$\sigma$. \textit{Right}: H$_2$ (2.12\,$\mu$m) line PV diagram tracing the molecular jet. Contours are from $3\sigma$ to $20\sigma$ in steps of $5\sigma$. Zoom-in panels of knots D (upper) and F (lower) with contours from $3\sigma$ to $80\sigma$ in steps of $20\sigma$ and from $3\sigma$ to $20\sigma$ in steps of $5\sigma$, respectively. Radial velocities are in the LSR frame.}
\label{fig:PVD_SLIT2}%
\end{figure*}

\begin{figure*}[ht]
\centering
\includegraphics[width=1.00\textwidth]{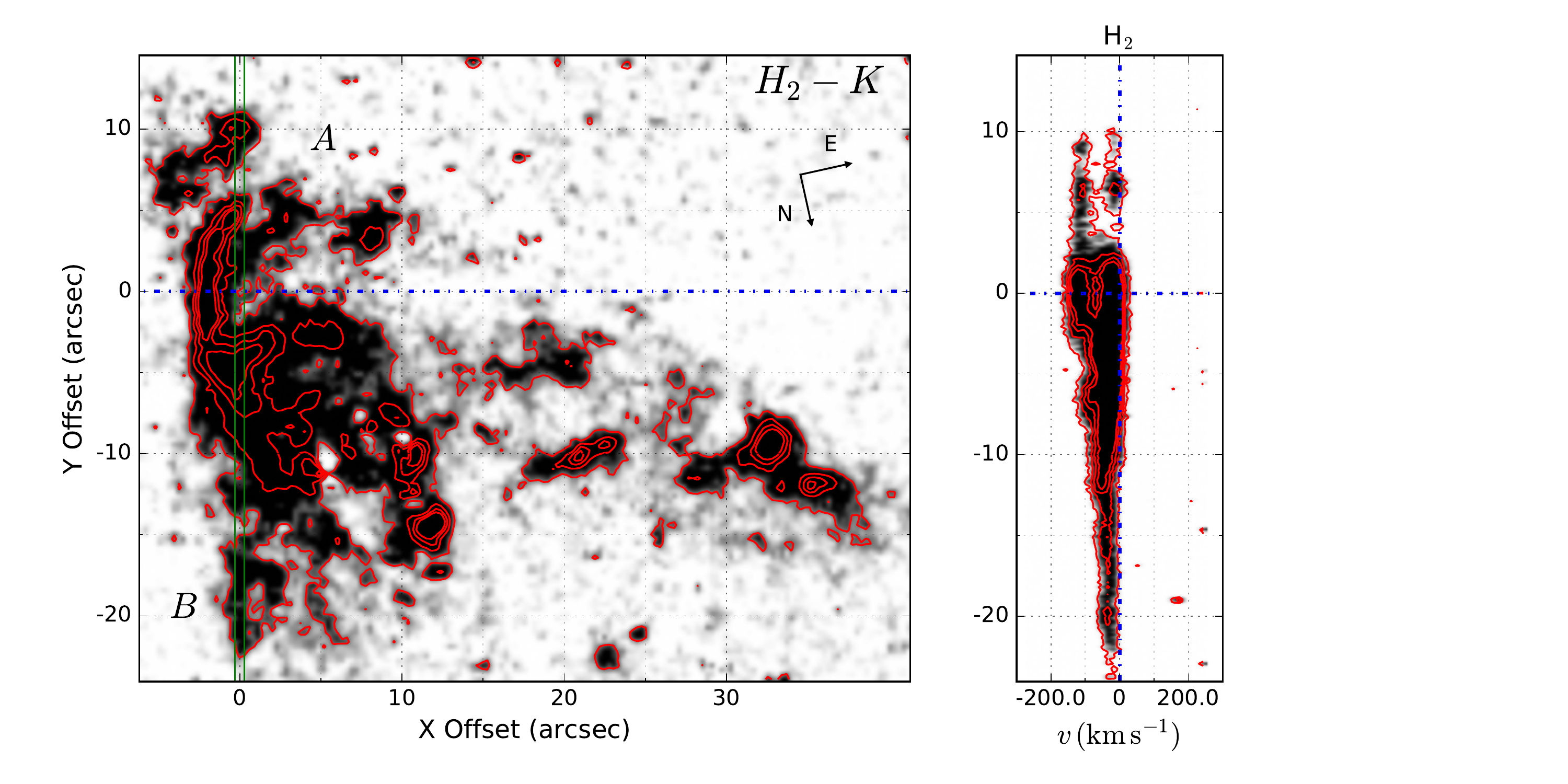}
\caption{PV diagram of slit 3. \textit{Left}: SOFI H$_2-K$ continuum subtracted of the blue bow shock. \textit{Right}: H$_2$ (2.12\,$\mu$m) line showing the two velocity components of the bow shock. Contours are from $3\sigma$ to $20\sigma$ in steps of $5\sigma$. Radial velocities are with respect to LSR.}
\label{fig:PVD_SLIT3}%
\end{figure*}

H$_2$ $1-0$\,S(1) emission is detected along the whole jet, as well as on the leading bow shock (knots A and B), whereas a faint [\ion{Fe}{ii}] emission is detected on knots D and F along the jet (Fig. \ref{fig:PVD_SLIT2}). No Br$\gamma$ emission is detected along the parsec-scale jet, nor is it associated with the leading bow shock (Figs. \ref{fig:PVD_SLIT2} and \ref{fig:PVD_SLIT3}). In addition to the $1-0$\,S(1) transition, other H$_2$ lines (i.e. $1-0$\,S(10) 1.6665 $\mu$m, $2-1$\,S(3) 2.0734 $\mu$m, and $2-1$\,S(2) 2.1542 $\mu$m) are detected in knot D and in the leading bow shock.

\begin{figure}[ht]
\centering
\includegraphics[width=0.5\textwidth]{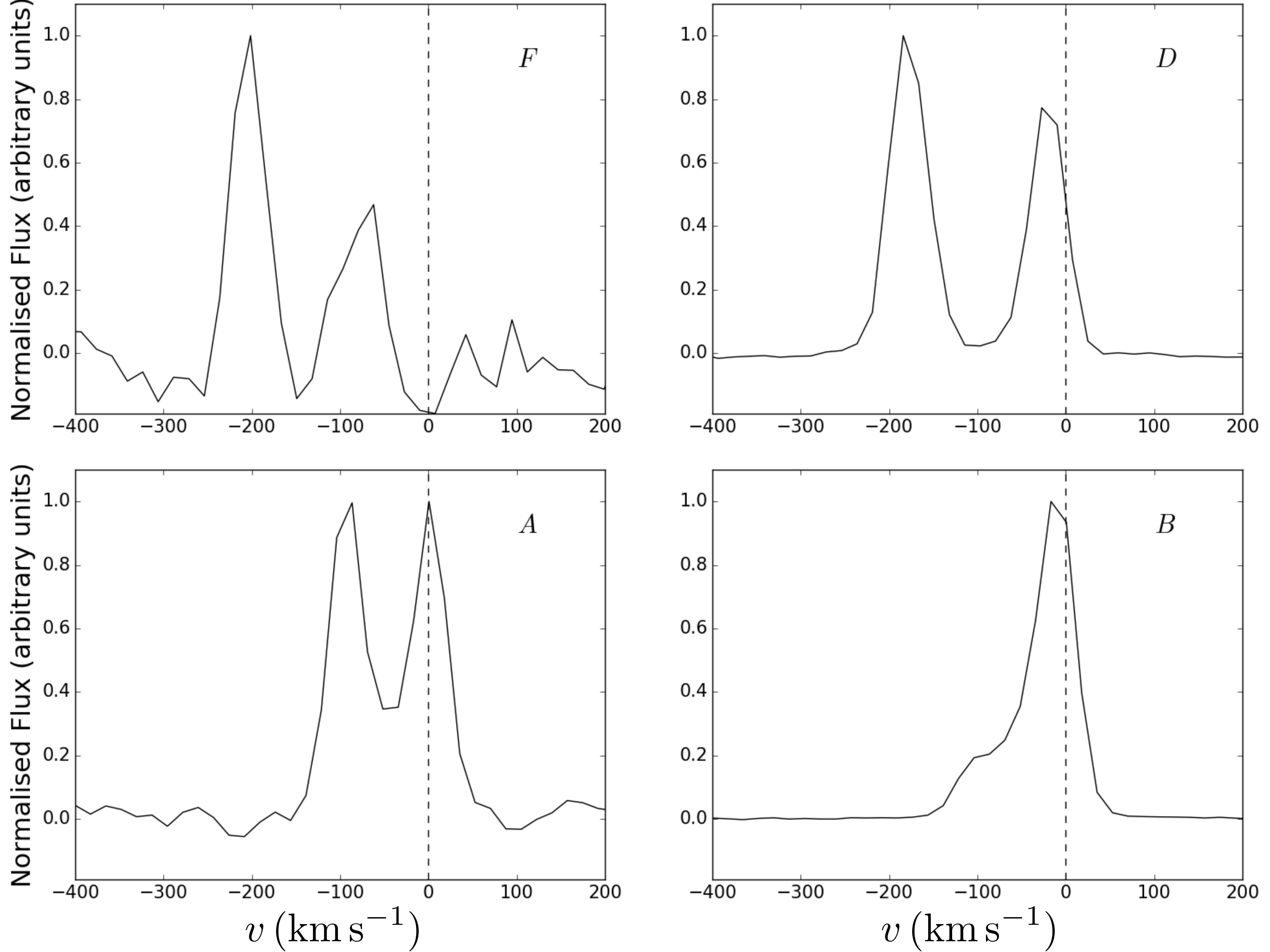}
\caption{H$_2$ 1-0 S(1) line profiles for the investigated knots. \textit{Top left}: F knot, \textit{top right}: D knot, \textit{bottom left}: A knot, and \textit{bottom right}: B knot. Radial velocities are with respect to the LSR.}
\label{fig:line_profile_H2}%
    \end{figure}

Figure \ref{fig:PVD_SLIT2} shows the position-velocity (PV) diagram for slit 2 encompassing the source and the jet, namely knots F and D. In the left panel, the SOFI H$_2$ continuum-subtracted image of the knots \citep[taken from][]{caratti2015} is shown as a reference. We note that the H$_2$ knots at positive x-offsets extending to the south-east might belong to another outflow or be part of the outflow cavity \citep{caratti2015}. The central and right panels of the figure display the spectral images of the [\ion{Fe}{ii}] and H$_2$ $1-0$\,S(1) lines, respectively. In both images the YSO continuum has been subtracted (at Y=0). Additionally, in the [\ion{Fe}{ii}] image an OH sky line was also subtracted, as evidenced by the residual noise at positive velocities. In this PV diagram, both atomic and molecular emission extend up to 70$\arcsec$ (or $\sim1$\,pc), which corresponds to the tip of knot D. Along the F and D knots, the [\ion{Fe}{ii}] radial velocities range from $\sim-60\mathrm{\,km\,s^{-1}}$ up to $\sim-200\mathrm{\,km\,s^{-1}}$. On source, the H$_2$ radial velocity is close to $\sim0\mathrm{\,km\,s^{-1}}$. Away from the source position (from 5$''$ on) the radial velocity ranges from $\sim-20\mathrm{\,km\,s^{-1}}$ to $\sim-200\mathrm{\,km\,s^{-1}}$. In particular, for knot D in [\ion{Fe}{ii}] and H$_2$, we identify a high-velocity component (HVC) and a low-velocity component (LVC) (see Fig. \ref{fig:PVD_SLIT2} right panel for the spectral image and Fig. \ref{fig:line_profile_H2} for the line profiles top panels). Additionally, we identify in the [\ion{Fe}{ii}] spectral image a high- and low-velocity pair at $2''$. The corresponding H$_2$ emission is not detected, most likely because this position is so close to the source that H$_2$ would be dissociated.

The H$_2$ emission close to the source (i.e. from $\sim5''$ to $25''$, knot F) is indicative of jet emission, with each velocity component originating in a separate sub-structure (see Fig. \ref{fig:PVD_SLIT2}, right panel). We note that the corresponding H$_2$ line profile (upper left panel of Fig. \ref{fig:line_profile_H2}) has been integrated over the whole knot F. In knot D, the H$_2$ PV diagram clearly shows two velocity components at roughly the same spatial location, suggesting a bow shock structure \citep[see e.g.][]{hartigan1987}, where the HVC is more likely to have a jet origin and the LVC an ambient material origin. However, the SOFI image of knot D does not show a clear bow shock morphology (Fig. \ref{fig:PVD_SLIT2}, left panel). Observations at higher spatial resolution are required to fully disentangle the structure.

The PV diagram for slit 3 is shown in Figure \ref{fig:PVD_SLIT3}, where the slit was placed along the bow shock, encompassing both wing (knot B) and head (knot A). The left panel shows the SOFI H$_2$ continuum-subtracted image of the blue-shifted bow shock (see Fig. \ref{fig:data_reduction}). In the right panel, the emission of the H$_2$ $1-0$\,S(1) line is shown. Inspection of Figure \ref{fig:data_reduction} reveals that A corresponds to the head of the bow shock and B the wing in this precessing jet. This interpretation is justified by the spectral image, which shows that A is mainly traced by the high-velocity component and B is mainly traced by the low-velocity component. The radial velocity ranges from a $\mathrm{\,few \,km\,s^{-1}}$ to $\sim-100\mathrm{\,km\,s^{-1}}$. Two velocity components are identified at some positions for each of the H$_2$ transitions (see e.g. Fig. \ref{fig:PVD_SLIT3} right panel and Fig. \ref{fig:line_profile_H2} bottom panels), and are indicative of a bow shock structure.

In order to calculate dynamic properties, we must first determine the jet radial velocity. In the case of knot A+B and D, the two H$_2$ velocity components are likely due to a bow shock geometry. Therefore, the radial velocity ($v_r$) reported in Table~\ref{tab:kinematics} refers to the HVC. Meanwhile, in the case of knot F, an average radial velocity along the knot was considered. This is because the jet mass to be used in our calculation relies on lower resolution observations, which do not resolve the sub-structure \citep{caratti2015}.
Moreover, due to the disc geometry of this object \citep[$i\sim45\degr$;][]{kraus2010}, the radial and tangential velocities of the jet can be assumed equal (i.e. $v_r=v_t$). This fact allows us to infer both tangential and total velocities. By combining the velocities with distance, length, and mass of each knot \citep[by fitting several H$_2$ transition intensities in a ro-vibrational diagram, based on low-resolution spectral data][]{caratti2015}, we can compute the main jet dynamic properties, reported in Table \ref{tab:kinematics}.

In particular, the mass-loss rate can be expressed as $\dot{M}_\mathrm{ejec}=M_\mathrm{knot} v_t/l$, where $M_\mathrm{knot}$ is the mass of the knot, $v_t$ is the tangential velocity, and $l$ is the length of the knot, which is given by $l=D\cdot\alpha$, where $D$ is the distance to the HMYSO and $\alpha$ is the extension of the knot on the sky. Values of the order of $\sim10^{-4}\mathrm{\,M_\odot\,yr^{-1}}$ were found for all the knots, within the error, suggesting that the mass ejection rate has been roughly constant, within a factor of two, in the history of this source (see Table \ref{tab:kinematics}, column 4 for individual values of the various knots). A rough estimate of the mass-loss rate for the atomic component using the [\ion{Fe}{ii}] 1.644 $\mu$m line was made for the F knot (highest signal for this line, S/N$\sim6.5$). Using the same reasoning as \citet{hartigan1994} (see their Eq. 5-10) and using the atomic values and cosmic abundances for iron \citep{mendoza1983,nussbaumer1980}, an expression for the mass-loss rate based on the iron line was derived. The expression can be written as 
\begin{eqnarray}
\dot{M}_\mathrm{[\ion{Fe}{ii}]}&=&1.9\times10^{-2}\left(\frac{N_e}{10^3(\mathrm{\,cm^{-3}})}\right)^{-1}\left(\frac{L}{L_\odot}\right)\frac{v_\perp}{l_\perp}\nonumber\\\nonumber\\&=&4.6\times10^{-6}\left(\frac{N_e}{10^3(\mathrm{\,cm^{-3}})}\right)^{-1}\mathrm{\,M_\odot\,yr^{-1}}
,\end{eqnarray}
where $N_e$ is the electron density, $L=0.04406\,L_\odot$ is the luminosity of the knot \citep[after being dereddened using][]{rieke1985}, $v_\perp=200\mathrm{\,km\,s^{-1}}$ is the velocity of the knot, and $l_\perp=2.5''$ is the length of the knot. Assuming typical electron density values from protostellar jets ($N_e\sim5\times10^4\mathrm{\,cm^{-3}}$), a mass-loss rate for the atomic component of the order of $\sim10^{-7}\mathrm{\,M_\odot\,yr^{-1}}$ is obtained.

On the other hand, the jet momentum is $P=M_\mathrm{knot}v_\mathrm{total}$, where $v_\mathrm{total}^2=v_r^2+v_t^2=2v_r^2$ is the total velocity. Momentum values of 86, 44, and 80 $\mathrm{M_\odot\,km\,s^{-1}}$ for the F, D, and A+B knots were found, respectively (see Table \ref{tab:kinematics}, column 5).

From the previous values, momentum rates can be also inferred, namely $\dot{P}=\dot{M}_\mathrm{ejec}v_\mathrm{total}$. Values of the momentum rate do not significantly vary along the jet, being of the order of $10^{-2}\mathrm{\,M_\odot\,yr^{-1}\,km\,s^{-1}}$ (Table \ref{tab:kinematics}, column 6).

Inferred kinetic energies of the knots ($K=1/2M_\mathrm{knot}v_\mathrm{total}^2$) range from $1.1\times10^{47}$ to $1.8\times10^{47}$\,erg (see Table \ref{tab:kinematics}, column 7).
Finally, we also derive the dynamical time of each knot, that is, $\tau_\mathrm{dyn}=d/v_t$, where $d$ is the distance of the knot from the central source. The dynamical time of the farthest bow shock provides us with an upper limit on the timescale of the ejecta, and, in turn, with an upper limit on the age of the central source. Values of the dynamical time range from 1800 to 26800\,yr for the closest and farthest emission, respectively (see Table \ref{tab:kinematics}, column 8).

\begin{table*}[ht]
\caption{Kinematic and dynamic properties of the IRAS\,13481-6124 parsec-scale jet derived from the H$_2$ $1-0$\,S(1) line.}
\label{tab:kinematics}
\begin{center}

\begin{tabular}{cccccccc}
\hline
\hline
\noalign{\smallskip}
Knot & $v_r$ & $M_\mathrm{knot}$ & $\dot{M}_\mathrm{ejec}$ & $P$  & $\dot{P}$ & $K$ & $\tau_\mathrm{dyn}$\\
 & ($\mathrm{km\,s^{-1}}$) & ($\mathrm{M_\odot}$) & ($10^{-4}\mathrm{\,M_\odot\,yr^{-1}}$) & ($\mathrm{M_\odot\,km\,s^{-1}}$) & ($10^{-2}\mathrm{\,M_\odot\,yr^{-1}\,\mathrm{km\,s^{-1}}}$) & ($10^{47}$ erg) & (yr)\\
\noalign{\smallskip}
\hline
\noalign{\smallskip}
A+B & -95$\pm$5 & $0.59\pm0.12$ & 1.5$\pm$0.3 & 80$\pm$14 & 2.0$\pm$0.3 & 1.1$\pm$0.2 & 26800$\pm$550\\
D & -180$\pm$5 & $0.17\pm0.05$ & 1.5$\pm$0.4 & 44$\pm$11 & 3.8$\pm$0.9 & 1.1$\pm$0.3 & 5400$\pm$50\\
F & $-145\pm7$ & $0.42\pm0.10$ & $2.7\pm0.7$ & $86\pm17$ & $5.6\pm1.1$ & $1.8\pm0.4$ & $1800\pm60$\\
\noalign{\smallskip}
\hline
\end{tabular}
\end{center}
\end{table*}

\section{LTE model of the Brackett emitting region} \label{sect:lte_model}

To model the physical conditions of the \ion{H}{i} emission detected on source in our SINFONI data, we used all the line fluxes from the Brackett series (Br26 to Br$\gamma$, see Table \ref{tab:lines}). Therefore, the main atomic gas parameters (close to the source), such as density, optical depth, and temperature can be derived. Homogeneous and isothermal conditions were assumed in the LTE model. Moreover, under LTE conditions, the kinetic temperature (Maxwell distribution) and excitation temperature (Boltzmann distribution) can be considered equal, that is, $T_{kin}=T_{ex}$. Due to the high density and high temperature conditions expected, the LTE assumption is plausible for the system. One expects LTE to be valid for electron densities higher than $\Sigma A_{ul}/\Sigma C_{ul}$, where $A_{ul}$ is the spontaneous radiation rates and $C_{ul}$ is the collisional de-excitation rates from level $u$ to level $l$ and the sum is over lower levels, that is, when collisions are the main source of excitation. Case B recombination can only reproduce our data when we set the density as high as $n=10^8\mathrm{\,cm^{-3}}$, which means that the gas would be in LTE.

\begin{figure}[ht]
\centering
\includegraphics[width=0.5\textwidth]{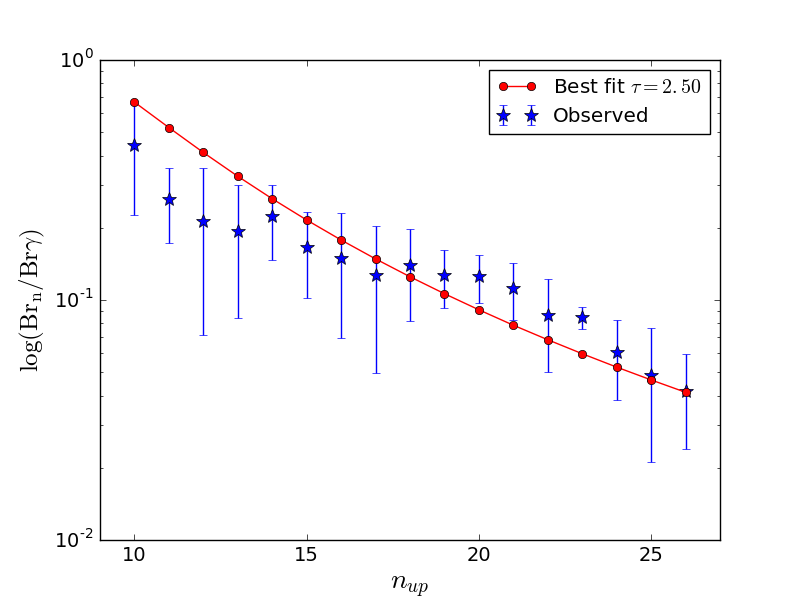}
\caption{Best fit of the observed Brackett decrement using our LTE model (red solid line). An optical depth of $\tau=2.5$ is inferred.}
\label{fig:brackett_decrement}%
    \end{figure}

In LTE conditions, the intensity can be written as 

\begin{equation}
I_\nu = B_\nu(T)\left(1-e^{-\tau(\nu,N_H,T)}\right),
\label{eq:intensity}
\end{equation}

\noindent where  $B_\nu$ is the Planck function and $\tau$ is the optical depth. $\tau(\nu,N_H,T)$ can be calculated assuming a hydrogen column density ($N_H$) and a temperature ($T$) for the system at a given frequency ($\nu$). The line ratios among Brackett lines can be used to infer physical conditions of the emitting gas and discern an approximative size of the emitting region. The Brackett decrement was computed using the ratio of the various Brackett lines with respect to the Br$\gamma$ line and assuming $T=10\,000$\,K. Figure \ref{fig:brackett_decrement} shows the observed Brackett decrement (blue stars) and the best fit (red dots) corresponding to an optical depth of $\tau_\mathrm{best}=2.5$ at $\lambda=2.1662\,\mu$m. The observed data were fitted using Eq.~\ref{eq:intensity}.

To estimate the size of the Br$\gamma$ emitting region, we used Eq.\,\ref{eq:radius} assuming, for simplicity, a spherical geometry:

\begin{equation}
r(T) = \left(\frac{4D^2F(\mathrm{Br}\gamma)}{B_\nu(T)\Delta\nu(1-e^{-\tau_\mathrm{best}})}\right)^{1/2},
\label{eq:radius}
\end{equation}

\noindent where $D=3.1$ kpc is the distance to the star, $F(\mathrm{Br}\gamma)=8.60\times10^{-12}\mathrm{\,erg\,s^{-1}\,cm^{-2}}$ is the integrated flux of the Br$\gamma$ line  \citep[corrected by extinction using the extinction law of][and assuming a value of $A_v=15$ mag from\citeauthor{caratti2015} \citeyear{caratti2015}]{rieke1985}  , and $\Delta \nu=\Delta v\cdot\nu/c$ is the line width, using the optical depth obtained from the best fit. A value of $\sim1$ au for the radius of the Br$\gamma$ emitting region was inferred. 

Using the estimated value of the radius of the Br$\gamma$ emitting region, one can derive the density 

\begin{equation}
n(R)=\frac{dM}{dt}\frac{1}{4\pi R^2v\mu m_H},
\label{eq:density}
\end{equation}

\noindent where $dM/dt=\dot{M}=1.8\times10^{-5}\,M_\odot\mathrm{\,yr^{-1}}$ is the mass-loss rate for the ionised jet \citep{purser2016}, $R=1$ au is the radius of the Br$\gamma$ region derived before, $v=500\mathrm{\,km\,s^{-1}}$ is the velocity of the jet close to the star \citep{purser2016}, $\mu=1.24$ is the mean atomic weight, and $m_H$ is the mass of the hydrogen atom. A value of $\sim2.5\times10^{9}\mathrm{\,cm^{-3}}$ for the density was found for the Br$\gamma$ emitting region.

\section{Discussion}\label{sect:discussion}

\subsection{Atomic au-scale jet} 

\label{sect:discussion_BrG}

Different approaches were used to study the innermost part of the parsec-scale jet of IRAS13481-6124. Examining the Br$\gamma$ on source, the LTE model of the Brackett emitting region, and spectro-astrometry allows us to study the jet down to mas scales. 

The P\,Cygni profile of the Br$\gamma$ indicates the presence of a powerful ionised wind at the base of the au-scale jet (Fig. \ref{fig:spectra} bottom right panel). As the other \ion{H}{i} line analysed, the line is clearly asymmetric, with strong self-absorption in the blue part of the line, produced by the outflow/wind, which reabsorbs the emission. The terminal radial velocity of the wind is around $-290\mathrm{\,km\,s^{-1}}$ (which corresponds to $v_\mathrm{total}\sim410\mathrm{\,km\,s^{-1}}$). The P\,Cygni profile is a peculiar feature that is not present in all HMYSOs. \citet{cooper2013} found only 13 YSOs (in a sample of 195 objects) that also display a P\,Cygni profile, probably due to the geometry of these objects. Usually, the Br$\gamma$ line in emission can have different origins: accretion of matter onto the star \citep{eisner2009,tambovtseva2014}, or ejected material in different manifestations -winds or jets- \citep{weigelt2011,stecklum2012,garcia-lopez2015,garcia-lopez2016,caratti2015HD98922}. See also \citet{coffey2010} and \citet{tambovtseva2016} for a detailed discussion of the origin of Br$\gamma$ emission. Nonetheless, one has to take into account that the broad profile of the \ion{H}{i} around the line peak can also be accounted for by the significant contribution expected from the disc rotation motions very close to the base of the jet ($v_\mathrm{kep}\sim130\mathrm{\,km\,yr^{-1}}$, for $M\sim20\,M_\odot$ and $R\sim10\,R_\odot$). It seems clear, however, for this particular object that Br$\gamma$ is tracing outflowing material and that the emission comes from the ionised jet \citep[see also][]{caratti2016}.

The optical depth derived from the LTE model ($\tau=2.5$) suggests that the Br$\gamma$ line is optically thick. Therefore, we can derive the density conditions of this region. A density of $\sim2.5\times10^{9}\mathrm{\,cm^{-3}}$ was found. In addition, a radius of $\sim1$\,au was inferred assuming a circular shape for the Br$\gamma$ emitting region. Using interferometric observations, \citet{caratti2016} found for the Br$\gamma$ emitting region an extension of $6.4-13$ au, observing a conical jet with an opening angle of $\sim30\degr$. If the projected circular region from the model is transformed into two projected cones with the same opening angle, an extension of $4.8$ au is found that matches the NIR interferometric observations
reasonably well. This shows that it is important to study the Br$\gamma$ line to constrain jet properties in YSOs, in particular of HMYSOs.

CRIRES spectro-astrometry shows a clear signature of well-collimated outflowing  material close to the star for the \ion{H}{i} Br$\gamma$ and Br$\alpha$ lines. The increasing photocentre offset with increasing velocity proves that the emission is coming from ejected material \citep{takami2001,whelan2005}. For the first time, spectro-astrometry was performed in the Br$\alpha$ line, which has similar properties (in size and orientation) as the Br$\gamma$ line. Position angles of $190\degr$ and $216\degr$ (and displacements of 11 and 15 au) were found for Br$\gamma$ and Br$\alpha$, respectively, suggesting that both lines might be tracing analogous regions. The good agreement between our results with those found by \citet{caratti2016} using interferometry on the Br$\gamma$ line is quite remarkable. In both studies, the size of the line-emitting region is almost identical. The PA found by \citet{caratti2015} for the parsec-scale jet ($\sim206\degr$) is consistent with the PA found in this study for the au-scale jet ($\sim190-216\degr$). The difference between the position angles might be due to the jet precession angle \citep[$\sim8\degr$,][]{caratti2015}. The consistency in size and position angle between the spectro-astrometry and the interferometry confirms the jet origin of the Br$\gamma$ emission.

Summarising, these observations strongly suggest that the Brackett lines are tracing the jet and its outflowing material. On the one hand, the Brackett emission cannot be explained as magnetospheric accretion (i.e. extension of the magnetosphere). Nor can the \ion{H}{i} emission originate from a Keplerian disc, as the spectro-astrometric results would not show a clear alignment with the jet axis and increasing velocity with increasing offset, but rather a displacement following the disc and with Keplerian rotation where the red and blue emission would be clearly differentiated with faster velocities closer to the star. On the other hand, \ion{H}{i} emission from UV pumping seems apparent, as we clearly detect lines (namely \ion{C}{i}, \ion{Mg}{i}, etc.), that seem to be connected to a photodissociation region (PDR). However, the Brackett line profiles show broad wings at high velocities. Moreover, one would expect a spherical shape in the distribution of PDR emission. We do not discard that some contribution of the Brackett lines may come from a PDR, but this seems negligible in comparison with the shocked emission. This can be shown by calculating the Str\"{o}ngrem radius for IRAS13481-6124 \citep[][]{drainebook}:

\begin{equation}
R_{SO}\equiv\left(\frac{3Q_0}{4\pi n_{H}^{2}\alpha_B}\right)^{1/3}=9.77\times10^{18}Q_{0,49}^{1/3}n_{2}^{-2/3}T_{4}^{0.28} \mathrm{\,\,cm},
\end{equation}

\noindent where $Q_{0,49}\equiv Q_0/10^{49}\mathrm{\,s^{-1}}$, $Q_0$ is the rate of emission of hydrogen-ionising photons, $n_2\equiv n_H/10^2\mathrm{\,cm^{-3}}$, $T_4=T/10^4$\,K, and $\alpha_B$ is given by

\begin{equation}
\alpha_B=2.54\times10^{-13}Z^2(T_4/Z^2)^{-0.8163-0.0208\ln(T_4/Z)} \mathrm{\,cm^3\,s^{-1}}
\label{eq:alphab}
\end{equation}

\noindent and $Q_0$ is given by

\begin{equation}
Q_0=7.58\times^{26}S_\nu D^2 \nu_{9}^{0.118}T_{4}^{-0.4933-0.0208\ln T_4} \mathrm{\,s^{-1}},
\end{equation}

\noindent where $S_\nu$ is the radio flux, $D$ distance to the object, and $\nu_9=\nu/\mathrm{GHz}$ is the frequency of the radio flux.

The $Q_0$ value was calculated using the radio flux on source from \citet{purser2016} ($9.1\pm0.07$\,mJy at 17.0\,GHz). Considering  $\log Q_0=44.75$, $T=10\,000$\,K, and $n=2.5\times10^{9}\mathrm{\,cm^{-3}}$ (from the LTE model, see Sect. \ref{sect:lte_model}) a value of $\sim0.25$\,au for the Str\"{o}mgren radius was inferred. The Str\"{o}mgren radius for this star is much smaller than the extension of the Brackett lines found using interferometry \citep[6.4-13\,au,][]{caratti2016}, confirming that the \ion{H}{i} Brackett lines are most likely tracing the au-scale jet.

\subsection{Molecular parsec-scale jet in context}

As was seen in the previous section, atomic hydrogen lines mainly trace the au-scale base of the jet. The composition of the NIR jet changes with distance from the source, with a smooth transition from atomic to molecular. At parsec scales, the observed jet becomes fully molecular, in the form of molecular hydrogen (H$_2$). Close to the central engine, the molecular hydrogen cannot survive since UV photons would dissociate the molecule and only atomic hydrogen would survive. Therefore, there is a transition with distance between the atomic and the molecular component. From our observations, the molecular component starts at $\sim5''-10''$ (i.e., $15\,000-30\,000$\,au) and extends up to $\sim170''-180''$ (i.e., $2.5-2.7$\,pc).

Notably, very high radial velocities are found in the H$_2$ emission lines along the knots of the parsec-scale jet: the F, D, and A+B knots. These values are well above the H$_2$ velocity dissociation \citep[$\sim50\mathrm{\,km\,s^{-1}}$,][]{smith1994}, which can reach up to $\sim80\mathrm{\,km\,s^{-1}}$ if a strong magnetic field is present \citep{lebourlot2002}. This phenomenon has previously been reported in other studies of jets driven by low-mass \citep{davis2000,chrysostomou2000} and high-mass protostars \citep{davis2004,caratti2008}. \citet{burton1986} proposed various mechanisms to explain the large line-widths observed in the H$_2$ lines, favouring the scenario that the shocked H$_2$ gas is in a medium that has been set in motion in fast-moving clumps. This scenario allows one to measure high-velocity H$_2$ components \citep{chrysostomou2000}, as the measured velocity is a combination of the medium and shock velocities.
\citet{davis2004} discussed that a combination of magnetically
mediated C-type shock and inclination of the flow with respect to the plane of the sky, H$_2$ line widths of up to $80-100\mathrm{\,km\,s^{-1}}$ are plausible. In this context, the radial velocity found in terminal bow shock (A+B knots; $\sim95\mathrm{\,km\,s^{-1}}$) can be easily explained. The leading bow shock must have moved in a steady medium and likely accelerated it as it propagated. This view is supported by the increasing velocity measured towards the central engine. The radial velocity found in knot D ($\sim100\mathrm{\,km\,s^{-1}}$) is faster than the leading A+B bow shock. The gas observed in this knot is probably moving into a medium that has been set in motion before by the forward shocks. In addition, F is the fastest knot found along the parsec-scale jet ($\sim200\mathrm{\,km\,s^{-1}}$), supporting the idea that this emission is embedded in an accelerated medium. \citet{devine1997} and \citet{chrysostomou2000} also reported a decrease in velocity with distance from the central source, which supports our results. This interpretation is also in agreement with the values of $\dot{P,}$ which slightly increase towards the central engine (see Fig. \ref{fig:m_dot_p_dot} bottom panel).

The dynamical ages of the A+B knots are around $\sim26800$\,yr, and they are the farthest structure in the system of the blue-shifted NIR jet. This value is consistent with the jet phase timescale of $\sim4\times10^4$\,yr suggested by \citet{guzman2012}. The dynamical age found for this knot is compatible, and might be even more accurate for the age of the star, with the age derived from SED modelling \citep[$\sim6\times10^4$\,yr,][]{grave2009}. Another interesting result concerns the roughly constant mass-loss rate value ($\sim10^{-4}\mathrm{\,M_\odot\,yr^{-1}}$) found along the various knots (see Fig. \ref{fig:m_dot_p_dot} top panel, and Table \ref{tab:kinematics}). This suggests that the ejection, and in turn, accretion of material has been constant in the formation history of IRAS\,13481-6124. The ejection-accretion processes are indeed closely related \citep[see e.g.][]{cabrit2007,cabrit2009}. For lower-mass YSOs, the ratios between the mass ejection rate ($\dot{M}_\mathrm{ejec}$) and the mass-accretion rate ($\dot{M}_\mathrm{acc}$) are found to be $\gtrsim0.1$ \citep{antoniucci2008}. If one considers a similar relation for the high-mass regime \citep[$0.1\lesssim\dot{M}_\mathrm{ejec}/\dot{M}_\mathrm{acc}\lesssim0.3$,][]{cabrit2007}, one obtains a mass-accretion rate of $3.33\times10^{-4}\lesssim \dot{M}_\mathrm{acc}\lesssim10^{-3}\mathrm{\,M_\odot\,yr^{-1}}$. Then, one can estimate the mass of the central object multiplying the mass-accretion rate by the dynamical age giving rise to $12\lesssim M_*\lesssim35\mathrm{\,M_\odot}$, consistent with the mass derived from SED modelling \citep[$\sim20\mathrm{\,M_\odot}$,][]{grave2009}.
Therefore it seems plausible that $\dot{M}_\mathrm{ejec}/\dot{M}_\mathrm{acc}$ ratio for at least this HMYSO is similar to those found in low-mass YSOs.

\subsection{Comparison between the NIR and the radio jet}

Our knowledge about HMYSOs is largely determined by radio observations, and fundamental dynamic properties, such as the mass-loss rate or mass-accretion rate, are based on radio measurements where only the ionised component is considered \citep[see e.g.][]{guzman2012,rosero2016,sanna2016,purser2016}. Physical properties derived from radio should be compared with those obtained from NIR observations for a better understanding of the formation of massive stars. To discern whether there is significant difference, we compared the dynamical properties of the radio jet \citep{purser2016} with those of the NIR jet.

We are in the position to estimate the ionisation fraction ($x_0$) of the HMYSO jet. \citet{purser2016} give an ionised mass-loss rate of $\dot{M}_\mathrm{ejec}=1.792\pm1.338\times10^{-5}\mathrm{\,M_\odot\,yr^{-1}}$ assuming a jet velocity of $500\mathrm{\,km\,s^{-1}}$ and an ionisation fraction of $0.2$. If we remove the velocity and ionisation fraction dependence from their Eq. 5, we can write $x_0\dot{M}_\mathrm{ejec}/v_\mathrm{jet}=7.168\pm5.353\times10^{-9}\mathrm{\,M_\odot\,yr^{-1}\,(km\,s^{-1})^{-1}}$ \citep[see][]{cesaroni2018}. From our NIR observations we can assume that the mass-loss rate close to the star is similar to the one derived along the parsec-scale jet ($\dot{M}_\mathrm{ejec}\sim1.766\pm0.276\times10^{-4}\mathrm{\,M_\odot\,yr^{-1}}$, roughly constant in the formation history of the star). Meanwhile, the PCygni profile of the Br$\gamma$ gives a range of velocities for the wind at the base of the jet close to the central source, that is, $v_\mathrm{jet}=400-1100\mathrm{\,km\,s^{-1}}$. Combining the above results, we obtain $x_0\sim2-5\%\pm3\%$. Therefore, a conservative upper limit of $x_0\lesssim8\%$ can be considered for the ionisation fraction of IRAS\,13481-6124, suggesting that only a small portion of the whole jet is ionised. Notably, similar results have also been found in both high-mass \citep{cesaroni2018} and low-mass \citep{ainsworth2013} regime. This calculation should be considered tentative because of the underlying assumptions and large uncertainties.

\begin{figure}[ht]
\centering
\includegraphics[width=0.5\textwidth]{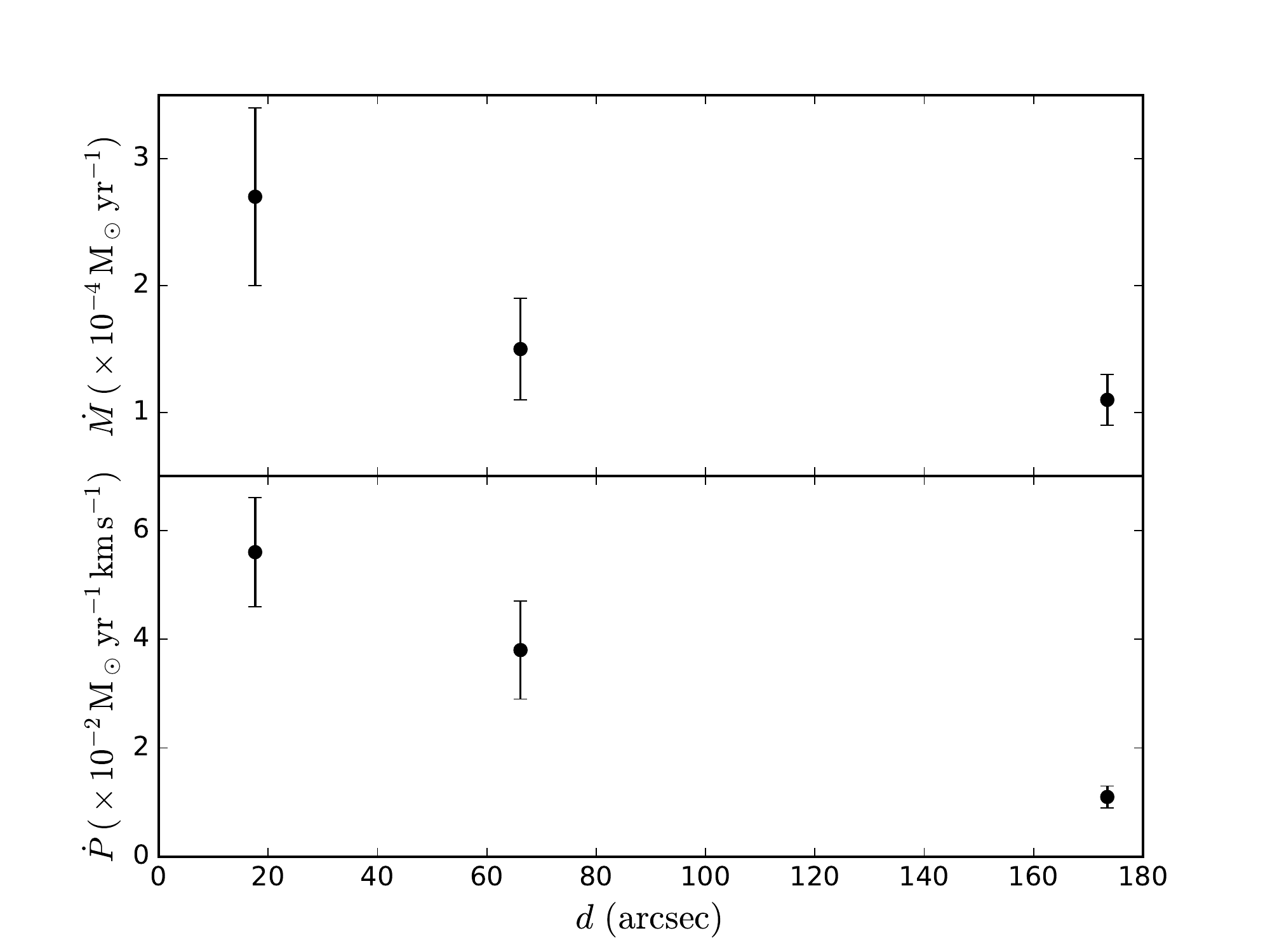}
\caption{\textit{Top panel}: Inferred mass-loss rates versus distance for the various knots. \textit{Bottom panel}: Inferred thrust versus distance.}
\label{fig:m_dot_p_dot}%
    \end{figure}

It is also worth pointing out that the low mass-loss rates of HMYSO radio jets seem to be a frequent characteristic in all ionised jets. For example, \citet{sanna2016} find a mass-loss rate for the radio jet of $\sim8\times10^{-6}\mathrm{\,M_\odot\,yr^{-1}}$ in a 20\,M$_\sun$ protostar, and \citet{purser2016} find a typical mass-loss rate of $\sim1.4\times10^{-5}\mathrm{\,M_\odot\,yr^{-1}}$ for a sample of HMYSOs. This issue has previously been considered by \citet{guzman2012}. To explain the discrepancy between the momentum of the radio jet and the molecular outflow, the authors suggested that the jet is not entirely ionised, as we are now demonstrating. 
Moreover, if one estimates the mass-accretion rate from the mass ejection rate of the radio jets (by assuming $0.1\lesssim\dot{M}_\mathrm{ejec}/\dot{M}_\mathrm{acc}\lesssim0.3$), one obtains a very low mass for the central object (assuming a typical age of few $10^4\,\mathrm{yr}$). In particular, for IRAS\,13481-6124, we would derive a value of $\dot{M}_\mathrm{acc}\lesssim 10^{-4}\mathrm{\,M_\odot\,yr^{-1}}$, which yields $M_*\sim3.5-6\,\mathrm{M_\sun}$ , in complete disagreement with the estimate of the central mass ($M_*\sim20\,\mathrm{M_\sun}$). Therefore, we conclude that the NIR jet is tracing the majority of the ejecta and the radio jet just a small portion of it. This is truly important because our mass-loss rate estimates for HMYSOs rely on radio measurements that trace a small percentage of the whole ejection. Nevertheless, both are likely part of the primary jet, as is revealed by comparing the thrust ($\dot{P}$) for the radio \citep[$\sim10^{-2}\mathrm{\,M_\odot\,yr^{-1}\,km\,s^{-1}}$][]{purser2016} and the NIR (roughly constant to $\sim10^{-2}\mathrm{\,M_\odot\,yr^{-1}\,km\,s^{-1}}$, see Fig. \ref{fig:m_dot_p_dot} bottom panel and Table \ref{tab:kinematics}) regimes.

\section{Conclusions} \label{sect:conclusions}

We used three ESO-VLT instruments, SINFONI, CRIRES, and ISAAC, to study the au- and parsec- scale jet as well as the immediate environment of the HMYSO IRAS\,13481-6124. Kinematic and dynamic properties were investigated at au and parsec scales, including the connection between the two. A comparison between the NIR jet and the radio jet was made. We summarise our main results and conclusions in the following points:

\begin{itemize}

\item Several emission lines were detected on source that are mainly associated with accretion and ejection activity (\ion{H}{i}, [\ion{Fe}{ii}]) with the presence of the disc ([\ion{Na}{i}]), and with UV-pumped emission (\ion{Fe}{ii}, \ion{C}{i}, \ion{K}{i}, \ion{Mg}{i}).

\item The Br$\gamma$ line was detected on source with a terminal radial velocity of $\sim-290\mathrm{\,km\,s^{-1}}$. From all three instruments, the characteristic P\,Cygni profile was identified, suggesting that the line is tracing ejection from a powerful bipolar wind very close to the central engine.

\item The technique of spectro-astrometry was applied to the Br$\gamma$ line and for the first time to the Br$\alpha$ line, revealing the atomic nature of the au-scale jet. The PA ($\sim190\pm15\degr$ and $\sim216\pm5\degr$ for Br$\gamma$ and Br$\alpha$, respectively) and the high collimation of the au-scale jet match the parsec-scale jet quite well. The photocentre offset with respect to the continuum of the Br$\gamma$ and Br$\alpha$ emitting region is at least $\sim11$\,au and $\sim15$\,au, respectively.

\item Molecular hydrogen (H$_2$) emission lines were used to derive dynamic and kinematic properties of the parsec-scale jet. Radial velocities were measured, and mass-loss rate, momentum, thrust, kinetic energy, and dynamical time were computed for the various knots that form the jet. Roughly constant mass-loss rates of the order of $\sim10^{-4}\mathrm{\,M_\odot\,yr^{-1}}$ were found along the parsec-scale jet. From this value, a mass-accretion rate of $\sim3\times10^{-4}-10^{-3}\mathrm{\,M_\odot\,yr^{-1}}$ was inferred. High H$_2$ radial velocities (from 100 to 200$\mathrm{\,km\,s^{-1}}$) were found likely due to the relative motion of the jet in an already-moving medium.

\item The ionisation fraction of the HMYSO jet driven was determined to have a tentative upper limit of $x_0\lesssim8\%$, suggesting that the radio jet  traces only a small fraction of the entire ejecta, whereas the NIR jet traces the majority.

\end{itemize}

In conclusion, the HMYSO jet of IRAS\,13481-6124 is traced mainly by atomic species at au-scales but molecular species at parsec-scales. Our derived ionisation fraction implies that the NIR component traces the bulk of the ejecta.

\begin{acknowledgements}
We would like to acknowledge Simon Purser for his inestimable help in gently transferring his radio data for the object IRAS13481-6124, which improved the discussion of the paper. We thank the anonymous referee for constructive comments. R.F. acknowledges support from Science Foundation Ireland (grant 13/ERC/12907). A.C.G. and T.P.R. have received funding from the European Research Council (ERC) under the European Union's Horizon 2020 research and innovation programme (grant agreement No.\ 743029). R.G.L has received funding from the European Union's Horizon 2020 research and innovation programme under the Marie Sk\l{}odowska-Curie Grant (agreement No.\ 706320). S.K. acknowledges support from an STFC Rutherford Fellowship (ST/J004030/1) and ERC Starting Grant (Grant Agreement No.\ 639889). 
\end{acknowledgements}


\bibliography{phd_bibliography}{}
\bibliographystyle{aa}

\end{document}